\documentclass[useAMS,usenatbib]{mn2e}

\usepackage{amssymb,natbib,graphicx,subfigure,color}

\newcommand{\be}{\begin{equation}}
\newcommand{\ee}{\end{equation}}
\newcommand{\bea}{\begin{eqnarray}}
\newcommand{\eea}{\end{eqnarray}}
\newcommand{\equ}[1]{equation (\ref{e:#1})}

\newcommand{\Equ}[1]{Equation (\ref{e:#1})}

\newcommand{\Fig}[1]{Fig.~\ref{f:#1}}

\newcommand{\ifm}[1]{\relax\ifmmode#1\else$\mathsurround=0pt #1$\fi}

\newcommand{\sggg}[1]{\textcolor{green}{[]}}

\def\log{{\rm\thinspace log}}

\def\pc{{\rm\thinspace pc}}
\def\kpc{{\rm\thinspace kpc}}
\def\Mpc{{\rm\thinspace Mpc}}
\newcommand{\hMpc}{\,\ifm{h^{-1}}{\rm Mpc}}

\def\Msun{\hbox{$\rm\thinspace M_{\odot}$}}
       
\def\yr{{\rm\thinspace yr}}

\def\Msunpc2{{\Msun\pc}^{-2}}
\def\Msunyrkpc2{{\Msun\yr^{-1}\kpc}^{-2}}
\def\magarcsec2{{\rm\thinspace mag\thinspace arcsec}^{-2}}

\newcommand{\apj}{ApJ}

\newcommand{\apjs}{ApJS}
\newcommand{\aj}{AJ}
\newcommand{\mnras}{MNRAS}
\newcommand{\aap}{A\&A}

\newcommand{\araa}{ARA\&A}
\newcommand{\nat}{Nature}

\def\MC{Monte Carlo{ }}
\def\VF{velocity field{ }}
\def\NMC{N_{\rm MC}}
\def\AREPO{{\small AREPO}}

\title[Tracer particles]{Following the flow: tracer particles in astrophysical fluid simulations}

\author[Genel, S. et al.]
{\parbox{20cm}{
Shy Genel$^{1}$\thanks{E-mail: sgenel@cfa.harvard.edu}, Mark Vogelsberger$^{1}$, Dylan Nelson$^{1}$, Debora Sijacki$^{1,2}$,\\ Volker Springel$^{3,4}$ and Lars Hernquist$^{1}$}\vspace{0.3cm}\\
$^{1}$Harvard-Smithsonian Center for Astrophysics, 60 Garden Street, Cambridge, MA 02138, USA\\
$^{2}$Institute of Astronomy and Kavli Institute for Cosmology, Cambridge University, Madingley Road, Cambridge CB3 0HA, UK\\
$^{3}$Heidelberg Institute for Theoretical Studies, Schloss-Wolfsbrunnenweg 35, 69118 Heidelberg, Germany\\
$^{4}$Zentrum f{\"u}r Astronomie der Universit{\"a}t Heidelberg, ARI, M{\"o}nchhofstr. 12-14, 69120 Heidelberg, Germany}

\begin{document}

\maketitle

\label{firstpage}

\begin{abstract}
We present two numerical schemes for passive tracer particles in the hydrodynamical moving-mesh code \AREPO, and compare their performance for various problems, from simple setups to cosmological simulations. The purpose of tracer particles is to allow the flow to be followed in a Lagrangian way, tracing the evolution of the fluid with time, and allowing the thermodynamical history of individual fluid parcels to be recorded. We find that the commonly-used `\VF tracers', which are advected using the fluid velocity field, do not in general follow the mass flow correctly, and explain why this is the case. This method can result in orders-of-magnitude biases in simulations of driven turbulence and in cosmological simulations, rendering the \VF tracers inappropriate for following these flows. We then discuss a novel implementation of `\MC tracers', which are moved along with fluid cells, and are exchanged probabilistically between them following the mass flux. This method reproduces the mass distribution of the fluid correctly. The main limitation of this approach is that it is more diffusive than the fluid itself. Nonetheless, we show that this novel approach is more reliable than what has been employed previously and demonstrate that it is appropriate for following hydrodynamical flows in mesh-based codes. The \MC tracers can also naturally be transferred between fluid cells and other types of particles, such as stellar particles, so that the mass flow in cosmological simulations can be followed in its entirety.
\end{abstract}

\begin{keywords}
hydrodynamics --
turbulence --
methods: numerical --
methods: statistical --
galaxies: formation --
cosmology: theory
\end{keywords}

\section{Introduction}
\label{s:intro}
The equations of hydrodynamics are usually solved in astrophysical applications using either of two general approaches: particle-based Lagrangian-like schemes such as Smoothed Particle Hydrodynamics (SPH), or mesh-based Eulerian-like schemes such as Adaptive Mesh Refinement (AMR). There are various advantages and shortcomings of each approach in terms of the accuracy of the solutions and of computational complexity. In SPH, physical quantities, such as mass, energy or entropy, are discretized into Monte Carlo particles that sample the underlying fluid elements and are moved in the simulation volume according to the equations of motions (e.g.~\citealp{LucyL_77a,GingoldR_77a,MonaghanJ_92a,SpringelV_02a,MonaghanJ_05a,PriceD_12a,HopkinsP_13a}). These are integrated using derived physical fields, such as the density and pressure, that are estimated from the particles using a smoothing kernel, while the particles themselves do not mix and have well-defined trajectories. Therefore, Lagrangian-like SPH schemes have the apparent advantage that it is possible to follow fluid resolution elements in time and track the evolution of their properties. However, this comes at a price of not integrating the equations of motion correctly (e.g.~\citealp{AgertzO_06a,SijackiD_12a,VogelsbergerM_12a,TorreyP_12a}). The fact that the mass density is estimated based on the Monte Carlo sampling of particles means that the mass continuity equation is not directly integrated at all. In contrast, in Eulerian-like schemes the discretized quantity is the volume itself, and conserved physical quantities of the fluid, such as mass, momentum, and energy, are exchanged between volume elements, i.e.~cells, according to the equations of motion. In this case, the fluid is mixed on the resolution scale, and therefore the information on its past evolution is lost.

It is possible to overcome this limitation of mesh-based schemes by introducing `tracer' particles, which can be passively advected with the fluid flow, and thereby track its Lagrangian evolution. Tracer particles can also record local instantaneous thermodynamic, or other, properties of the fluid, providing information on the evolution of the fluid at any point in space and time. If required for a specific problem, they can also model some physical processes in a `sub-grid' fashion, and so actively modify e.g.~the dynamical or chemical evolution of the simulation.

There are several types of problems in which the ability to follow the evolution of Lagrangian fluid elements is crucial. In particular, one of the most debated issues in galaxy formation is the question of `how galaxies get their gas', i.e.~what is the hydrodynamical, kinematical, and thermal evolution of gas as it accretes from the intergalactic medium into dark matter haloes and eventually settles in galaxies possibly forming stars (e.g.~\citealp{WhiteS_78a,KeresD_05a,DekelA_09a}). Clearly, the ability to follow gas back in time that has settled into galaxies in a Lagrangian manner is necessarily required to quantitatively investigate this question (e.g.~\citealp{PichonC_11a,DuboisY_12a,TillsonH_13a,NelsonD_13a}). A second example is the study of mixing in various turbulent environments, such as in the interstellar or intergalactic media, where it is useful to know for different `final' phase-space coordinates what is the mixture of `initial' phase-space coordinates of the fluid (e.g.~\citealp{FederrathC_08a,MitchellN_09a,VazzaF_11a,VazzaF_11b,ZavalaJ_12a}). In addition, time-dependent physical processes such as reaction networks for molecular hydrogen, dust grains, cosmic rays, or nuclear burning in supernova explosions, may depend on the past evolution of a Lagrangian resolution element. Thus, knowledge of the Lagrangian history of the fluid can be required not only for post-processing analysis, but for the modeling itself (e.g.~\citealp{NagatakiS_97a,EnsslinT_02a,BrownE_05a,TravaglioC_04a,FederrathC_08a,SilviaD_10a}).

Here we report on the implementation and characterization of two tracer particle schemes in the moving-mesh code \AREPO{ }\citep{SpringelV_10a}. The code employs a quasi-Lagrangian scheme, in the sense that the movement of the mesh closely follows the hydrodynamical flow. Nevertheless, the employed `finite-volume' scheme is not entirely Lagrangian, since fluid can be exchanged between cells, as in the more common use of static meshes. Mass exchange between cells is required because the cell shapes cannot in detail account for any arbitrary mass flow. Therefore, to trace Lagrangian mass elements with \AREPO, we implemented tracer particle schemes.

In particular, we implemented two entirely independent tracer particle schemes. We first present a scheme in which tracer particles are implemented as massless particles that are advected in space using the local velocity field \citep{HarlowF_65a}. This approach is routinely used in the literature, e.g.~in conjunction with various numerical techniques for direct numerical simulation (DNS) of turbulent flows (e.g.~\citealp{YeungP_88a}), as well as for groundwater modeling (e.g.~\citealp{LuN_94a,AndersonM_02a}). It has also been implemented for astrophysical applications in several AMR codes, such as {\small ZEUS-3D} \citep{EnsslinT_02a}, {\small FLASH} \citep{FisherR_08a,DubeyA_12a}, {\small ENZO} \citep{BryanG_13a}, and {\small RAMSES} \citep{PichonC_11a,DuboisY_12a}. We find that the general result of this approach, particularly evident in certain situations that we identify herein, is a failure to follow the flow of the underlying fluid. This makes this scheme unreliable for studying mass flows in cosmological simulations of galaxy formation. We therefore introduce a fundamentally different approach that is based on a \MC sampling of fluid motions. In this scheme, tracer particles are not represented by phase-space coordinates, but are rather attached to resolution elements, such as fluid cells or stellar particles, and are carried with them. Whenever two cells/particles exchange mass, they also exchange the appropriate fraction of their tracer particles. In this way, the tracer particles follow the mass flow accurately by construction, with limitations due to \MC statistical noise and increased spatial diffusion.

This paper is organized as follows. In Section \ref{s:implementation} we present the two schemes and give a detailed description of their implementation in \AREPO. In Section \ref{s:simple_tests} we present a set of simple setups that demonstrate the workings of the two methods, and most importantly, their limitations. In particular, Section \ref{s:sine_wave} shows that the \VF tracers do not in general follow the mass flow correctly, and explains the origin of this behaviour. In Section \ref{s:complex_tests} we further compare the two methods with more complex, astrophysically-motivated, setups: driven isothermal turbulence (Section \ref{s:turbulence}) and cosmological simulations of galaxy formation (Section \ref{s:cosmo}). In Section \ref{s:cluster} we apply the \MC tracers to study the thermodynamical history of the atmosphere of the Santa Barbara Cluster. Finally, in Section \ref{s:summary} we summarize our results and conclude.

\section{Implementation}
\label{s:implementation}
\subsection{Velocity field tracers}
\label{s:vel_tr}
Velocity field tracers are massless particles that do not affect the dynamical evolution of the simulation, but are advected with the flow by using an estimate for the velocity field of the fluid at the positions of the particles. There exist various methods for the calculation of the local fluid velocity based on the computational hydrodynamical mesh, and for the time integration of tracer particle positions based on the estimated local velocities \citep{AndersonM_02a}. For example, the velocity assigned to the tracer particles can be simply that of the nearest cell, but it is more common to use a Cloud-In-Cell (CIC) interpolation scheme or yet higher order interpolations such as Triangular-Shaped Cloud. It has been found, however, that such variants do not significantly affect the results \citep{FederrathC_08a,VazzaF_10a,KonstandinL_12a}. Using these velocities, tracer particle positions can then be updated with an Euler method, based on the hydrodynamical time step. Higher order time-integration schemes are also used in some applications (e.g.~\citealp{ArznerK_06a,PopovP_08a,SukH_09a}).

\begin{figure}
\centering
\includegraphics[width=0.475\textwidth]{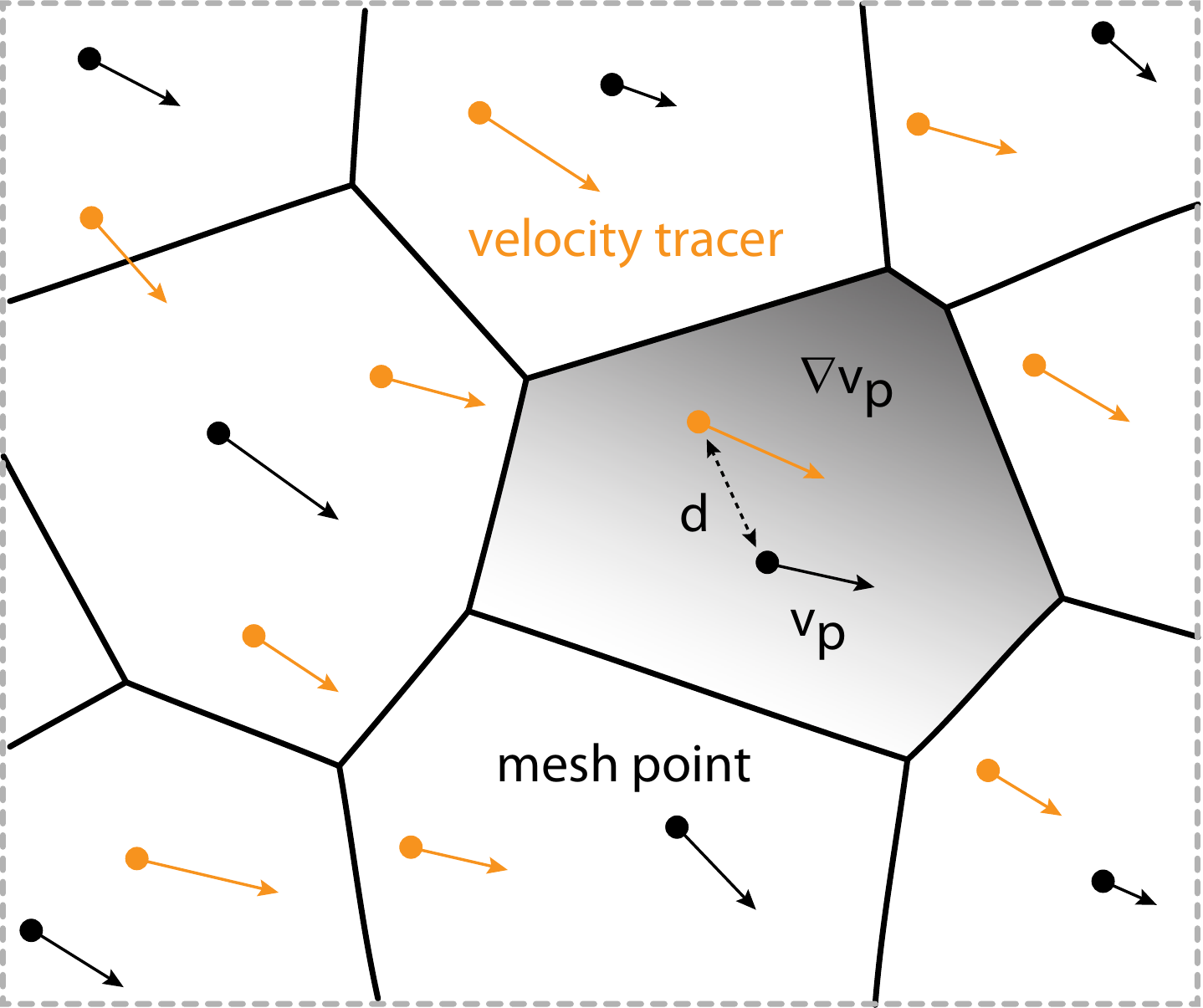}
\caption{Schematic representation of our \VF tracer particle implementation. Shown are cells with mesh-generating points ({\it black}) and tracers ({\it orange}). We note that the number of tracers per cell is arbitrary and depends on the details of the hydrodynamical flow. The black arrows denote the velocities of the fluid at the mesh-generating points. Orange arrows denote the velocities assigned to the individual tracer particles. These velocities in general differ from the fluid velocities at the mesh-generating points, since we exploit the linear velocity field gradient in each Voronoi cell (represented by the color gradient in the central cell) to interpolate the velocities at the tracer particle positions within the cell.}
\vspace{0.5cm}
\label{f:vel_tr_imp}
\end{figure}

In our implementation of a velocity field tracer particle scheme in \AREPO, each tracer particle moves according to the linearly interpolated velocity field of its parent cell; i.e.~the cell in which the tracer particle is embedded. We find the parent cell of each tracer particle by searching for the nearest mesh-generating point, exploiting the convexity of Voronoi cells. This is done at each time step for each active tracer particle using the neighbour tree, with an initial search radius guessed based on the distance to the nearest cell in the previous time step. Tracer particles are marked as active if their parent cell is active, and inherit the time steps from their parent cells, making their time integration adaptive based on the underlying cell hierarchy. We then use the velocity field gradient of the parent cell to interpolate the velocity field at the tracer position. The cell gradients are calculated based on the Green-Gauss theorem with a stencil that includes all adjacent cells, and are additionally constrained by a slope limiter that ensures that the linearly reconstructed velocities on face centroids do not exceed the maxima or minima among all adjacent cells (for more details, see \citealp{SpringelV_10a}). The gradient information is readily available for the cells, because it is already calculated for the MUSCL-Hancock steps in the finite volume solver. Once the velocity is assigned to each active tracer particle, we drift it with its individual time step. The tracer particle time integration is implemented with second-order accuracy consistently with the hydrodynamical integration of \AREPO. In this implementation, the primary computational overhead is due to the required cell lookup. \Fig{vel_tr_imp} shows a schematic representation of the velocity field tracer technique.

As we demonstrate below, the \VF tracer technique is not guaranteed to recover the correct mass flows in certain, quite general, hydrodynamical situations. This is, for example, the case for turbulent flows, and for gas accreted onto the centres of haloes; i.e.~common situations in simulations of galaxy formation. We note that this effect can be clearly seen in some previous studies that used \VF tracer particle implementations \citep{VazzaF_10a,PichonC_11a,PriceD_10a}, however it was, to the best of our knowledge, never interpreted correctly as a serious failure in astrophysically-related studies.

\subsection{\MC tracers}
\label{s:MC_tr}
The most basic requirement for tracing the Lagrangian history of a fluid element is that the tracer particles accurately follow the fluid mass flow. To satisfy this requirement, we introduce a second, novel scheme in which tracer particles (which could be thought of more appropriately as `unique tags') are attached to particular resolution elements. Finite volume fluxes between neighbouring cells are already calculated during each active time step for the hydrodynamics. Tracer particles are then exchanged between neighboring cells according to these mass fluxes, where the fraction moved corresponds to the fraction of the cell mass transferred across the boundary face. Due to the finite number of such tracers, this results in a \MC sampling of the underlying fluid mass flux over the computational domain. As a result, and by construction, these tracers are forced to follow the fluid flow, and do not suffer from biases seen in the velocity field interpolation approach.

\begin{figure}
\centering
\includegraphics[width=0.475\textwidth]{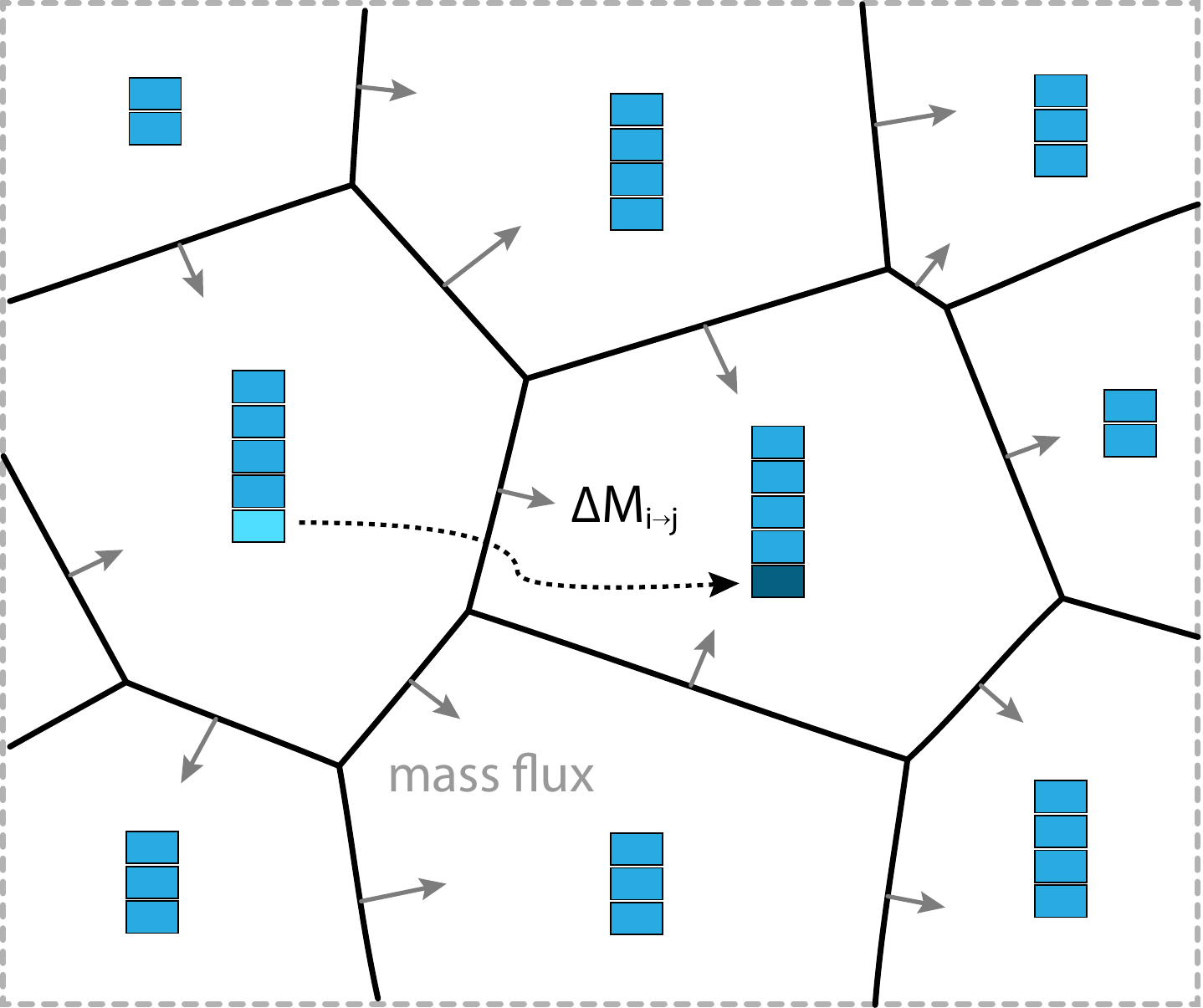}
\caption{Schematic representation of our \MC tracer implementation. The cell setup is the same as in \Fig{MC_tr_imp}. Tracers are collected in buckets for each cell and they have no individual phase-space properties, i.e.~they are just an additional property of each cell. Arrows indicate outgoing mass fluxes for each cell. Tracers are then probabilistically exchanged between cells based on these mass fluxes.}
\vspace{0.5cm}
\label{f:MC_tr_imp}
\end{figure}

This \MC sampling can be efficiently inlined in the finite volume solver loop, which runs over all Voronoi cell faces. This face list is constructed beforehand based on the volume Voronoi/Delaunay tessellation. During the face loop we keep track of two relevant quantities. First, the current list of each `original' tracer $\alpha$ of each cell $i$, which includes all the tracers that belong to cell $i$ before the finite volume solver exchange loop starts. Second, the `reduced' mass $\widetilde{M}_i$ of each cell $i$, which is initialized before the loop to be equal to the mass of cell $i$ ($M_i$), and updated during the loop only for outgoing fluxes, but not for incoming fluxes. When the face loop arrives to the face between cells $i$ and $j$ where there is mass flux $\Delta M_{i, j}$, we consider each `original' tracer $\alpha$ (i.e.~ignoring tracers that joined cell $i$ during the current face loop) that still belongs to cell $i$ (i.e.~ignoring tracers that were already moved from cell $i$ to another cell during the current face loop). The probability of each of those tracers to leave the cell and be moved to cell $j$ is then given by
\begin{equation}
p_{i, j}^{\rm flux} \, = \, \frac{\Delta M_{i, j}}{\widetilde{M}_i}.
\label{e:tracer_P_flux}
\end{equation}
To decide whether tracer $\alpha$ should leave cell $i$ we draw a random number $x_\alpha \in U(0,1)$. The tracer is moved to cell $j$ if $x_\alpha < p_{i, j}^{\rm flux}$. Finally, before continuing to the next face in the loop, we update the value of the reduced mass $\widetilde{M}_i$ to be $\widetilde{M}_i-\Delta M_{i, j}$, regardless of the actual tracer transfers that have or have not occurred. In this process, only outgoing mass fluxes $\Delta M_{i, j}$ from cell $i$ to cell $j$ are considered, while incoming fluxes into cell $i$ are automatically treated by the outgoing fluxes of its neighboring cells with which it has a common face.

In such a Monte Carlo-based implementation, tracers are not described by phase-space coordinates within the cell, i.e.~tracers contain no sub-resolution spatial information. This is different from the \VF approach, where gradient information within individual cells is taken into account by interpolating the fluid velocity field to the exact tracer particle position in the cell. In the \MC approach, tracers are completely mixed within the cell (but still keep their individual identities). We therefore keep them in a global and parallelized linked list, where each list entry has a tracer ID, tracked fluid properties and pointers to the next and previous tracers in the tracer list that belong to the same cell. Each fluid cell then needs only a pointer to the head of its own sub-list to find the tracers associated with it. Tracer exchanges between cells can then be implemented as efficient linked list operations on this global list.

In addition to finite volume solver fluxes, mesh refinement and derefinement operations must be explicitly handled in the \MC approach\footnote{Mesh refinement and derefinement operations do not directly affect the \VF tracers since they have their own spatial positions, and only read out the current fluid velocity field once they are active.}. In a refinement step, we split a cell $i$ into two cells, where each daughter cell is given some mass fraction of the parent cell. This is implemented in practice by reducing a mass $\Delta M_{i, j}$ from the parent cell $i$ that is then assigned to a new cell $j$. We sample this probabilistically in the same way we treated mass fluxes in the finite volume solver step, where the probability is now given by
\begin{equation}
p_{i, j}^{\rm refine} \, = \, \frac{\Delta M_{i, j}}{M_i}
\label{e:tracer_P_refine}
\end{equation}
for each tracer that belongs to the parent cell $i$. In a derefinement step we remove a mesh-generating point from the cell that should be derefined. The mass and other conserved quantities of the cell are then spread to the neighboring cells according to the volume fraction of the removed cell claimed by those cells. Since this is effectively the same operation as a mass flux $\Delta M_{i, j}$ from the cell $i$ that is being derefined to other cells $j$, we use \equ{tracer_P_flux} to distribute the tracers accordingly. For derefinement, when the loop arrives at the last neighboring cell $j_l$ there should be $p_{i, j_l}^{\rm flux}=1$ since 
\begin{equation}
\sum_j \Delta M_{i, j}=M_i.
\label{e:tracer_P_derefine_sum}
\end{equation}
However, round-off errors may introduce a small deviation, hence for the last neighboring cell we explicitly set $p_{i, j}^{\rm flux}=1$, such that all the remaining tracers are moved before cell $i$ is removed from the mesh.

The \MC tracer particle scheme can be naturally extended for applications that include additional particle types beyond pure hydrodynamics, such as astrophysical applications with stellar particles, or black hole (sink) particles. We allow not only tracer exchanges from gas to stars during star-formation, but also from stars to gas for stellar mass loss processes. When a gas cell is converted into a stellar particle, its tracers keep their individual identities, in particular their stored fluid quantities that represent their evolution history. When a stellar particle of mass $M_*$ is spawned out of a gas cell $i$ of mass $M_i$, each tracer belonging to cell $i$ is given a probability
\begin{equation}
p_{i, *}^{\rm SF} \, = \, \frac{M_*}{M_i}
\label{e:tracer_P_SF}
\end{equation}
to be moved to the new stellar particle, much like the case for mesh refinement. An analogous treatment is given to black hole particles. In addition, when a stellar particle releases mass into its neighboring gas cells (such as in models for stellar winds, see \citealp{VogelsbergerM_13a}), the procedure is again similar to the finite volume solver fluxes or derefinement cases. Specifically, during a loop where a stellar particle with mass $M_*$ releases mass to its neighboring gas cells, when some mass $\Delta M_{*, j}$ is transferred into gas cell $j$, each tracer particle that still belongs to the star particle is given a probability of 
\begin{equation}
p_{*, j}^{\rm recycling} \, = \, \frac{\Delta M_{*, j}}{\widetilde{M}_*}
\label{e:tracer_P_recycling}
\end{equation}
to be moved to cell $j$. To do that, we keep track of the reduced mass $\widetilde{M}_*$, such that it is initialized to $M_*$, and updated to be $\widetilde{M}_*-\Delta M_{*, j}$ after the exchange of mass and tracers with cell $j$ is done.

\subsection{Initialization of tracers}
\label{s:initialization}
At the beginning of the simulation we generate tracer particles based on the initial conditions of the fluid. For \VF tracer particles, we assign typically one tracer particle to each cell, which inherits the position of its parent cell; i.e.~its mesh-generating point. This is valid as long as the masses of the cells in the initial conditions are equal. To initialize more than one \VF tracer particle per cell, the underlying mass distribution has to be sampled correctly. This can be non-trivial for general initial conditions, such as cosmological ones. This task is simpler for initial conditions with a uniform (or piecewise-uniform) density, in which case the \VF tracers need not be initialized at the positions of the mesh-generating points, but rather can sample the volume with a finer, uniform, grid. For \MC tracers, we assign $\NMC\ge1$ tracers to each cell. The sampling noise is reduced by choosing a larger number of tracers per cell $\NMC$, as discussed in Section \ref{s:advection}.

\subsection{Recording tracer history}
\label{s:recording}
Both types of tracer particles can follow the evolution of the fluid along their trajectories by continuously recording properties of the fluid locally. To this end, we keep an array of properties for each tracer particle that is updated every active time step. We store, for example, the maximum fluid temperature a tracer encounters, the maximum Mach number as read from the Riemann solver, and the times at which these maximum values are reached. By zeroing those values after they are written to a snapshot file, we are able to keep the high time resolution throughout the simulation, as the history of the quantities is continuously recorded even if they drop after reaching their maximum values. The maximum gas temperature is required, for instance, to study cosmological gas accretion and distinguish cold and hot mode accretion in \AREPO{ }(or any Eulerian-like mesh code) in the same manner as was done in previous SPH simulations (\citealp{NelsonD_13a} vs.~\citealp{KeresD_05a}). For the \MC tracers, we also record quantities such as the number of cell exchanges they experienced, and the last time they were associated with a stellar particle. We implement the tracer particles with their recording capabilities in state-of-the-art simulations of galaxy formation including metal cooling, star-formation and stellar mass return, black hole evolution, and various feedback processes, described in \citet{VogelsbergerM_13a}, such that the mass exchange through all represented phases of baryonic matter is self-consistently followed.

\section{Simple tests: demonstrating the methods and their limitations}
\label{s:simple_tests}
\subsection{Advection}
\label{s:advection}
We begin by presenting the results of a very simple test, which is nevertheless quite instructive in providing basic insights into the workings of the two methods, in particular for the \MC tracers. We set up a two-dimensional $10\times10$ periodic box with uniform-density fluid ($\rho=1$) that has a uniform velocity in the positive $x$-direction $v_x=1$ (all units are in internal code units). We distribute $10\times10$ mesh-generating points in the initial conditions such that the initial grid is Cartesian and regular. Hence, the side length of each cell is $\Delta x=1$, and the cell-crossing time is $t_c=\Delta x/v_x=1$. In the initial conditions, one \VF tracer is placed at the position of each mesh-generating point, and $\NMC=10$ \MC tracers are initialized within each cell. The box is run to time $t_f=50$ in two ways, once with a static mesh, and once with a moving mesh.

\begin{figure*}
\centering
\includegraphics[width=1.0\textwidth]{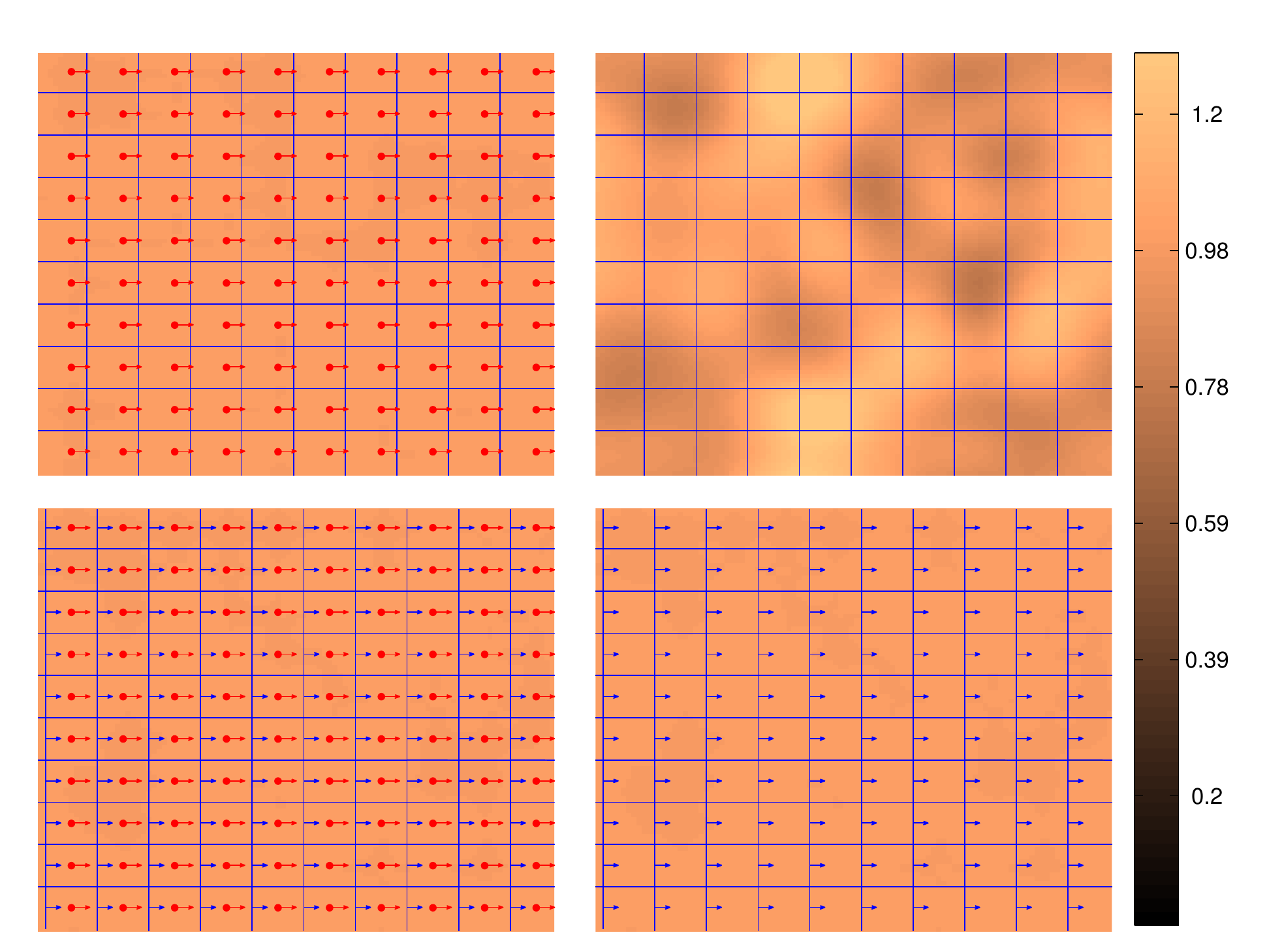}
\caption{The results of the uniform advection test at time $t=49.2$. In each panel, the background colour shows an SPH-like density, as indicated by the colour bar, and the grid shows the Voronoi mesh. The left panels show the density of the \VF tracers, with the red points showing their locations. The right panels show the density of the \MC tracers, which do not have their own positions, as they only 'belong' to gas cells. In the top panels the mesh is fixed, while in the bottom panels a moving mesh is used. In this simple test, only the \MC tracers with a static mesh show density deviations with respect to the uniform fluid density, a result of \MC sampling noise.}
\vspace{0.5cm}
\label{f:advection_maps}
\end{figure*}

\Fig{advection_maps} shows the results at time $t=49.2$. In each panel, the grid shows the Voronoi mesh, marked with blue arrows for the moving mesh runs (bottom panels). The background colour shows an SPH-like density estimate, as indicated by the colour bar. The left panels show the density of the \VF tracers, with the red points showing their locations, and the right panels show the density of the \MC tracers. The \VF tracers retain their original regular distribution, since they move with the local velocity (indicated by the red arrows), which is uniform in the box at all times in this setup. At $t=49.2$ they are displaced by $0.2$ cell lengths from their original positions in the centres of the cells, as expected. Their positions are independent of whether the mesh is moving or not, since the local fluid velocities are reconstructed identically in both cases. As a result, the \VF tracers follow the fluid perfectly in this test.

In contrast, the movement of the \MC tracers does depend on the movement of the mesh, since the \MC tracers are attached to gas cells and only move from one cell to another when there is mass exchange. In the moving mesh simulation there is no mass exchange between cells, since the mesh moves exactly with the flow thanks to the quasi-Lagrangian nature of \AREPO, which for this simple flow becomes fully Lagrangian. As a result, each cell keeps its original \MC tracers, and the \MC tracers then follow the mass exactly (lower right panel). However, when the mesh is static (upper right panel), the code is fully Eulerian, and the flow occurs solely due to mass exchanges between cells. In this case, the tracers are exchanged as well, but in a probabilistic way that introduces \MC statistical noise, as we show next.

\begin{figure}
\centering
\subfigure[]{
          \label{f:advection_Nt_distribution_Ni5}
          \includegraphics[width=0.475\textwidth]{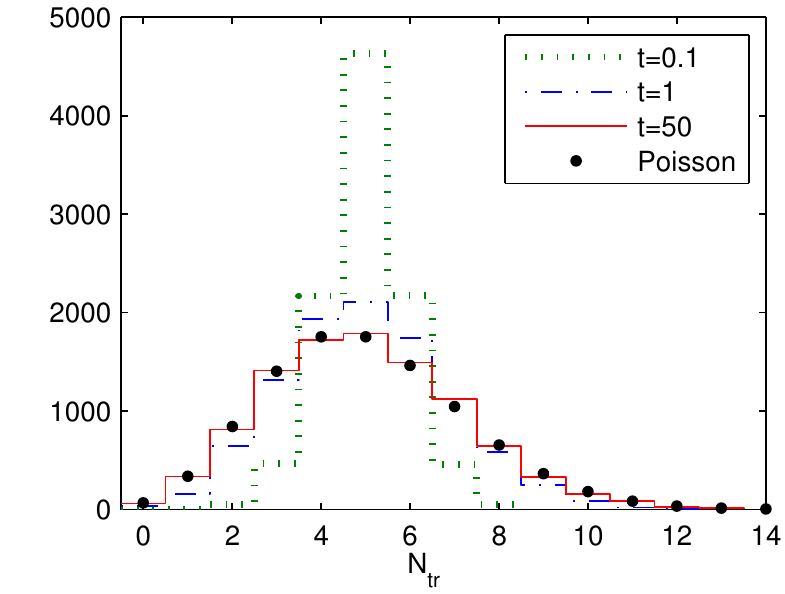}}
\subfigure[]{
          \label{f:advection_Nt_distribution_vs_Ni}
          \includegraphics[width=0.475\textwidth]{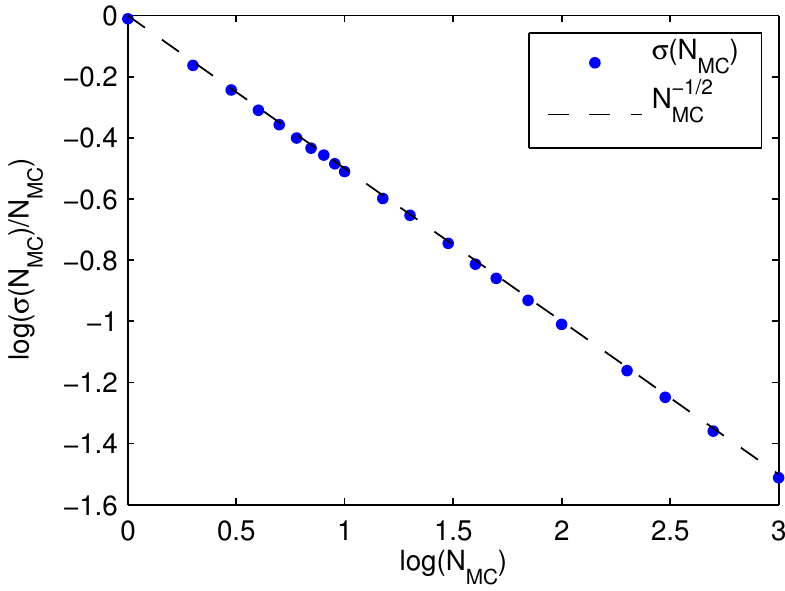}}
\caption{{\it Top:} The distribution of the number of \MC tracers per cell in the uniform advection test with $\NMC=5$, for three different times in units of the cell crossing time $t_c=\Delta x/v_x=1$. These distributions are compared with a Poisson distribution ({\it dots}), and it is demonstrated that the distribution approaches a Poisson distribution within roughly one cell crossing time $t_c$. {\it Bottom:} The standard deviation of the same distribution as in the top panel, at the final time $t=50t_c$, for different initial values $\NMC$ of \MC tracers per cell. The agreement with the expectation from a Poisson distribution $\sigma(\NMC)/\langle\NMC\rangle=\langle\NMC\rangle^{-1/2}$ is excellent. All these results are for simulations with a static mesh.}
\vspace{0.5cm}
\label{f:advection_Nt_distribution}
\end{figure}

By construction, the rate at which \MC tracers are added/removed from their cells, per unit mass, is constant, and such events are independent of one another. Therefore, the distribution of the number of \MC tracers per cell approaches a Poisson distribution within a typical cell-crossing time $t_c$. This is demonstrated in \Fig{advection_Nt_distribution} for a sized-up version of this test with $100\times100$ cells in a $100\times100$ box, with a static mesh. In \Fig{advection_Nt_distribution_Ni5}, where we use $\NMC=5$ \MC tracers per cell in the initial conditions, the distribution of the number of \MC tracers per cell is shown for different times and compared to a Poisson distribution with a parameter $\lambda=\NMC=5$, showing excellent agreement for $t\gtrsim t_c$. In \Fig{advection_Nt_distribution_vs_Ni}, the relative standard deviation of that distribution at $t=50t_c$ is plotted against $\NMC$, showing an excellent agreement with the expected property of the Poisson distribution $\sigma(\NMC)/\langle\NMC\rangle=\langle\NMC\rangle^{-1/2}$. This remains true in all the tests shown below in this paper --  the distribution of the number of \MC tracers (normalized by mass) follows a Poisson distribution. While the statistical nature of the \MC tracers is clearly disadvantageous compared to the \VF tracers for this simple test, as it introduces Poisson noise, we note that this is a minor concern for most, if not all, practical applications. First, \MC tracers consume significantly less computational resources than \VF tracers (see below), such that it is relatively easy to use large numbers of \MC tracers per cell, thereby reducing the noise, as demonstrated in \Fig{advection_Nt_distribution_vs_Ni}. Second, rarely will a single cell be of special interest. It is more meaningful to investigate the Lagrangian history of a group of cells that represents some physical entity, such as a galaxy, or a certain radius in a gaseous halo, for which the total \MC tracer number will typically be very large, and therefore the noise low. Hence, for practical applications, the \MC tracers reproduce the true mass distribution well even for a static mesh.

\begin{figure}
\centering
\subfigure[]{
          \label{f:advection_disp_distribution_vs_time}
          \includegraphics[width=0.475\textwidth]{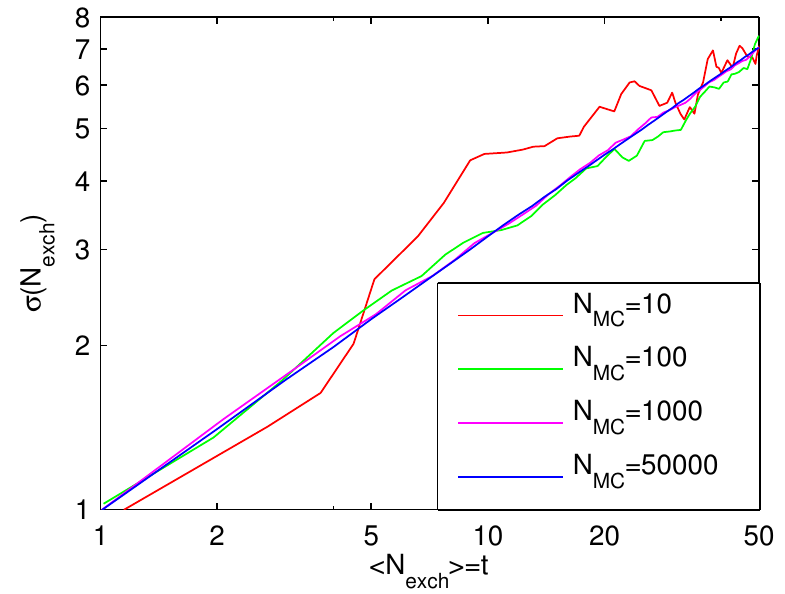}}
\subfigure[]{
          \label{f:advection_disp_distribution_vs_res}
          \includegraphics[width=0.475\textwidth]{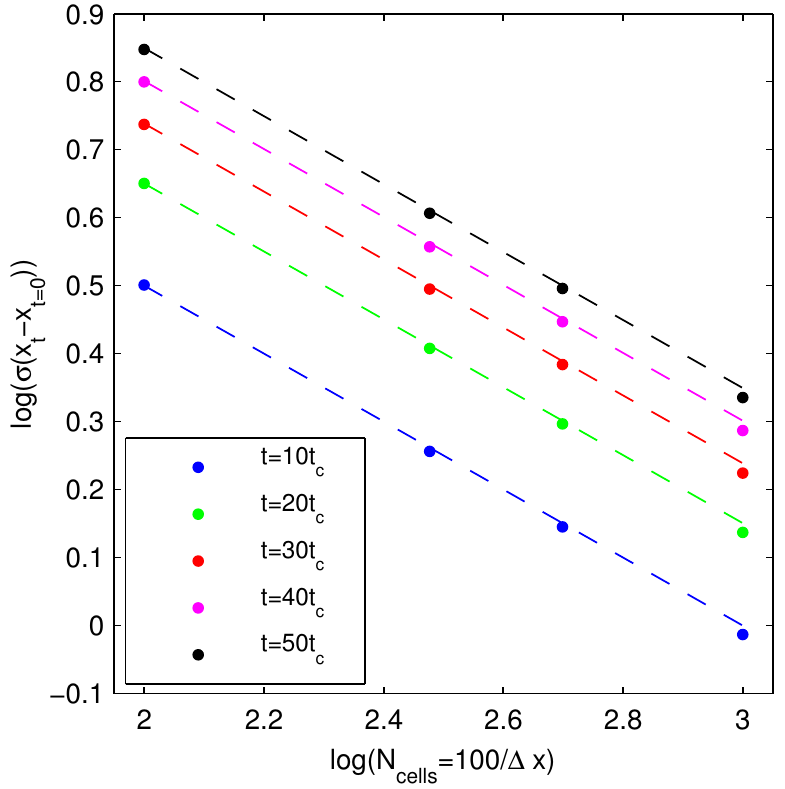}}
\caption{{\it Top:} The standard deviation of the distribution of number of cell exchanges $N_{\rm exch}$ that \MC tracers undergo as a function of time for the uniform advection test. The mean number of cell exchanges follows the fluid accurately, hence in this setup, with $v_x=\Delta x=1$ and a static mesh, $N_{\rm exch}=\frac{\int{v_x {\rm d}t}}{\Delta x}=t$. However, the distribution of $N_{\rm exch}$ is approximately a Poisson distribution, and $\sigma(N_{\rm exch})\approx\sqrt{\langle N_{\rm exch}\rangle}$. The different curves show this relation, \equ{Nexch_error}, for groups of randomly-selected \MC tracers of different numbers. For a large number of \MC tracers, \equ{Nexch_error} holds very precisely. {\it Bottom:} The standard deviation of the distribution of distances covered by different \MC tracers as a function of the resolution of the simulation $N_{\rm cells}=100/\Delta x$, for various times $t=Xt_c$, where $X=10,20,30,40,50$ (dots, from bottom to top). The dashed lines show \equ{distance_error}, $\sqrt{(x_{t=Xt_c}-x_{t=0})\Delta x}$. The excellent agreement demonstrates the validity of \equ{distance_error}. Both panels are symmetric with respect to time; i.e.~they show the exact same behavior if a group of tracers at a certain position is followed backward, rather than forward, in time.}
\vspace{0.5cm}
\label{f:advection_disp_distribution}
\end{figure}

However, there is a different kind of error that cannot be reduced by using a larger number of \MC tracers. The probabilistic nature of the motion of the tracers means that they propagate in a random walk with respect to the underlying fluid motion -- in other words, they diffuse with respect to the fluid. To see this, consider that for each \MC tracer, every simulation time step is a Bernoulli trial with a probability $p_{i, j}^{\rm flux}$ to move to another cell. For independent trials, which is generally a good approximation (but see below), this makes the distribution of $N_{\rm exch}$, the number of cell exchanges each \MC tracer experiences, follow the Poisson binomial distribution. In the typical case $p_{i, j}^{\rm flux}\ll1$, the Poisson binomial distribution has the property that the variance equals the mean, up to a small correction factor\footnote{For equal probabilities $p_{i, j}^{\rm flux}$ in all time steps, as in the uniform advection test discussed in this section, the Poisson binomial distribution becomes the binomial distribution, which for a large number of time steps and $p_{i, j}^{\rm flux}\ll1$ approaches the Poisson distribution.}, i.e.
\be
\label{e:Nexch_error}
\sigma(N_{\rm exch})=\sqrt{\langle N_{\rm exch}\rangle}.
\ee
The standard deviation of the distribution of number of exchanges (which for this particular test problem, with $\Delta x=v_x=1$, equals the distance propagated in the positive $x$-direction) is shown in \Fig{advection_disp_distribution_vs_time} versus the mean number of exchanges, or time, and confirms \equ{Nexch_error}. For a given problem, the fluid propagates a certain physical distance $D=D_{\rm mesh}+D_{\rm exch}$, which is composed of the movement of the mesh $D_{\rm mesh}$ and movement with respect to the mesh by fluxes between cells $D_{\rm exch}$. We can write $D_{\rm exch}=\Delta x N_{\rm exch}$, hence the statistical error of the distance $D$ travelled by any individual \MC tracer is typically
\bea
\label{e:distance_error}
\sigma(D)&=&\sigma(D_{\rm exch})=\Delta x\sigma(N_{\rm exch})\\\nonumber
&\approx&\Delta x\sqrt{N_{\rm exch}}=\sqrt{D_{\rm exch}\Delta x},
\eea
where the transition from the first to the second line uses \equ{Nexch_error}; i.e.~it relies on the assumption of independent probabilities $p_{i, j}^{\rm flux}$ in different time steps. We stress that \equ{distance_error} is not only a proportionality relation, but a true equality. Thus, the spread of the distance travelled by different \MC tracers becomes smaller with improving resolution. This is demonstrated in \Fig{advection_disp_distribution_vs_res}. \Equ{distance_error} is time-symmetric; i.e.~it represents both the spatial spread of the \MC tracers around the `true' coordinates at time $t_f$ of the fluid element they started with at time $t_i$, and the \MC tracers spread around the `true' original coordinates at time $t_i$ of a fluid element selected at a certain position at time $t_f$. The meaning of this type of error is that the Lagrangian history of the \MC tracers is not exact, but noisy. Whether or not this poses a serious problem depends on the simulation setup and the details of the flow. It is clear, however, that a moving mesh has a significant advantage for the performance of the \MC tracers thanks to its own quasi-Lagrangian behavior that significantly reduces $D_{\rm exch}$.

\begin{figure}
\centering
\includegraphics[width=0.475\textwidth]{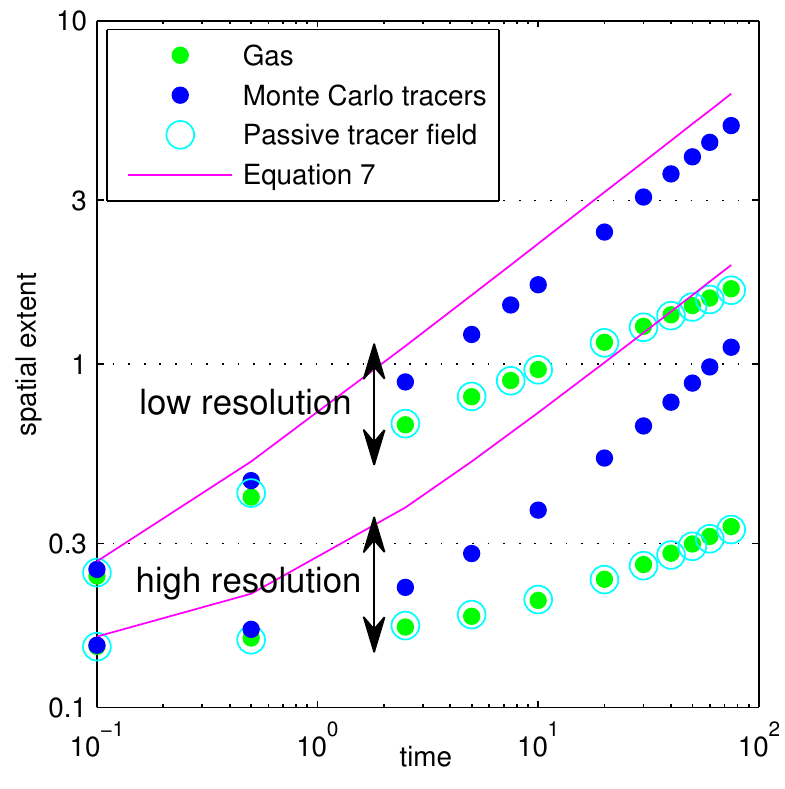}
\caption{The spatial spread of the fluid (green dots), \MC tracers (blue dots), and a passive tracer field (cyan circles), which were initially located in a thin over-dense region of a contact discontinuity advection test, as a function of time. The \MC tracers extend with time over larger spatial regions, as a result of their larger numerical diffusion compared with that of the fluid. Their spread does not follow exactly that given by \equ{distance_error} (magenta) since the cell exchanges are in general not independent, but the slope given by \equ{distance_error} is correct, and the normalization offset is of order unity.}
\vspace{0.5cm}
\label{f:CD_spread_vs_time}
\end{figure}

To see how the diffusion of the \MC tracers relates to the numerical diffusion of the fluid itself, we run a similar test, but in this case advect a contact discontinuity. We set up a box of side length $100$, where the region $0<x<0.5$ has a density of $13.4$, and at $0.5<x<100$ the density is $0.94$, i.e.~a mean density of $1$ in the box, and a density contrast of $14.3$ across the discontinuity. Additionally, we initialize the over-dense cells with a passive tracer field, which has zero value everywhere else in the box. The passive tracer field evolves as a dye cast on the fluid, its value for each cell indicating the fraction of the mass that was initially in the over-dense cells. The pressure is constant throughout the box, and the fluid is given a uniform velocity of $v_x=1$. \Fig{CD_spread_vs_time} shows the width (standard deviation) of the spatial extent of the contents of the initially over-dense region as a function of time. The fluid over-density (green) and the passive tracer (cyan) show excellent agreement, and their spread evolves approximately as $\propto t^{0.3}$. We run this test at two resolution levels, with $200$ and $2000$ cells in the $x$-direction, shown by the top and bottom sets of curves, respectively. All spreads are smaller for the higher resolution run. Specifically, the fluid spread is reduced by a factor $\approx5\approx(2000/200)^{0.7}$, which is consistent with the finding of \citet{SpringelV_10a} that the $L1$ error of a contact discontinuity advection test scales as $L1\propto N^{-0.75}$. 

The spread of the \MC tracers (blue) scales as $\propto t^{0.5}$, as expected from \equ{distance_error} (magenta). The normalization of the actual spread of the \MC tracers is lower than what \equ{distance_error} gives, as opposed to the case for the uniform density advection test (\Fig{advection_disp_distribution}). The reason for this normalization offset is that where the density is not constant, cell exchanges become somewhat correlated, such that $N_{\rm exch}$ does not exactly follow a Poisson binomial distribution anymore. Nevertheless, the scaling $\sigma(D)\propto\sqrt{D_{\rm exch}}$ from \equ{distance_error} still holds. The \MC tracer spread scales with resolution by a factor of $\approx4.5$, better than the factor $\sqrt{10}$ expected from \equ{distance_error}. This is interpreted as increased correlations between the exchanges, as in the higher resolution case the fluid over-density remains sharper, hence the density gradients larger, than in the lower resolution case.

Most importantly, \Fig{CD_spread_vs_time} demonstrates that the \MC tracers are more diffusive than the fluid, as the spread of the \MC tracers that resided initially in the over-dense region (blue) is higher than that of the fluid (green), and scales more strongly with time, or distance travelled. Without this enhanced diffusion, the \MC tracers would be exact \MC sampling points of the passive tracer field. To understand the origin of the enhanced diffusion, consider that each time step in \AREPO{ }can be considered as a sequence of three steps: reconstruction, evolution, and averaging \citep{SpringelV_10a}. The fluxes of \MC tracers follow those of the fluid correctly (evolution), and they are by construction mixed inside their cells (averaging). However, the \MC tracers undergo no reconstruction step. Even though the fluxes of the \MC tracers follow those of the fluid, which are calculated using a second-order piecewise linear scheme, the \MC tracers themselves are advected similarly to a first-order donor-cell scheme since they are completely mixed on the sub-grid level. This means, e.g.~for flux vectors that point in the positive direction, that a tracer particle that enters a cell from the `negative face' can already in the next time step leave the cell through its opposite `positive face'. In contrast, the sub-cell linear reconstruction of the density field can prevent the fluid from being immediately spread out over the entire cell and propagating to the opposite face\footnote{Following and preserving free boundaries is a generic problem in fluid dynamics, see e.g.~a discussion of the volume of fluid (FOV) and other methods in \citet{HirtC_81a}.}. This is possible despite that the fluid is mixed inside the cell during the averaging step, as the reconstruction comes after the averaging and before the next evolution. Therefore, the fluid has lower advection errors, and hence lower numerical diffusion than the tracers. This can in principle be improved upon by introducing some sub-grid model for the distribution of the \MC tracers inside the cells; e.g.~having the \MC tracers remember the direction from which they arrived to their current cell, and not allowing them to `overrun' other tracers. We leave such improvements to the algorithm for future work.

Finally, we note that the \VF tracers in the contact discontinuity advection test are perfectly advected and {\bf do not} diffuse at all with respect to their initial distribution, as the velocity is constant in the simulation box, just as in the case of the uniform density advection. In contrast, advection of the density jump relative to the grid will always involve some numerical diffusion. This is a first example for a case where the \VF tracers evolve differently from the fluid, a topic that is the main focus of the next section.

Our \MC tracers are significantly less costly in terms of computational resources than the \VF tracers. In our implementation, the \VF tracer particles are realized as a special type of collisionless particle, therefore internal memory is required to store their phase-space coordinates, masses (which are, however, all identically zero), IDs, and several other variables, depending on the nature of the simulation. In contrast, the \MC tracers are realized with a special linked list, such that each tracer keeps only information about its previous and next tracer in the list, and its own ID. Hence, if no physical quantities are recorded by any of the tracer types, the internal memory required per \VF tracer is roughly four times larger than what is required per \MC tracer, a value that depends on the nature of the simulation and the precision mode (single or double) used to store the different variables. Since their evolution does not involve neighbour searches, \MC tracers also require fewer computations. The exact computational load per tracer depends somewhat on the problem. For this uniform flow test, adding one velocity field tracer per cell adds $\approx3\%$ to the total calculation time, and each \MC tracer per cell adds only $0.7\%$.

\subsection{Adding velocity gradients: a converging/diverging flow}
\label{s:sine_wave}
In the simple test we presented in the previous section, it was demonstrated that the \VF tracers can follow the flow of the fluid very accurately. In that test, the \MC tracers follow the fluid correctly only on average, albeit with some statistical noise that in general depends on the problem and on `how Lagrangian' the motion of the mesh itself is. In this section, however, we show that once the flow becomes even slightly more complex, the \VF tracers cannot follow the fluid correctly, even in an average sense.

\begin{figure*}
\centering
\includegraphics[width=1.0\textwidth]{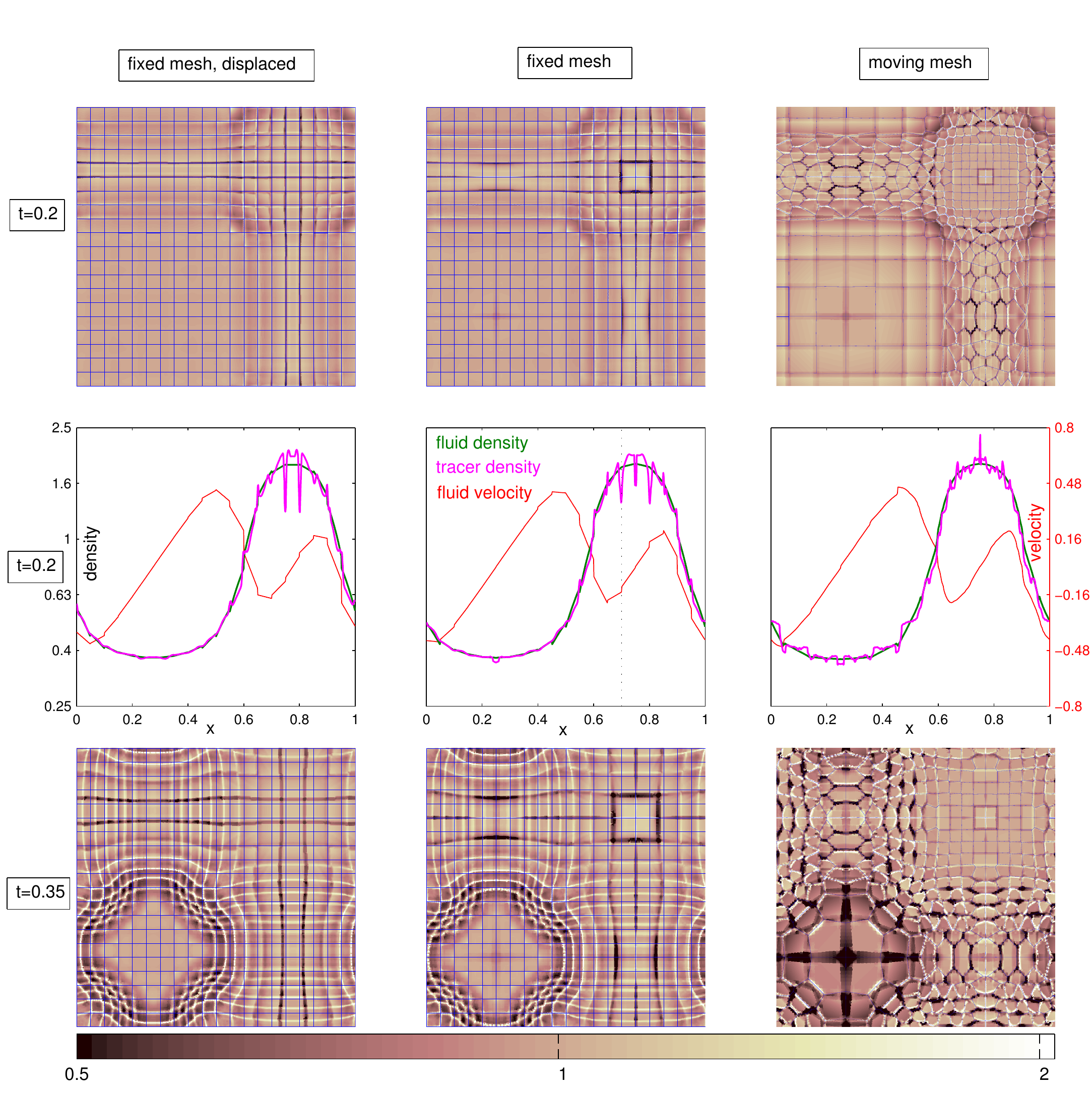}
\caption{The converging/diverging flow test at times $t=0.2$ (top and middle rows) and $t=0.35$ (bottom row), after discontinuities have appeared in the reconstructed velocity and density fields. The background colour in the top and bottom rows represents the density ratio of the \VF tracers with respect to the fluid, with the colour bar at the bottom indicating the scale, and blue lines delineating the Voronoi mesh. The panels in the middle row show one-dimensional profiles averaged along the $x$-axis of the fluid density (green), the \VF tracer density (magenta), and the velocity field in the $x$-direction (red). The middle column shows our default setup with a static mesh, while the side columns show the same physical setup with technical variations: on the right the system is evolved using a moving mesh, and on the left the initial velocity field was displaced by half a cell size ($\Delta_x=0.025$) with respect to the default setup, while the configuration of the (static) mesh has not been changed. It is evident from the top and middle panels that the \VF tracers do not follow the fluid correctly especially where discontinuities in the reconstructed fields appeared; i.e.~along certain cell interfaces (an example is indicated with a dotted line at $x=0.7$ in the middle panel). The bottom panels show that those over/under-densities are later advected and leave their formation sites along cell interfaces, thereby creating spurious sub-cell structures. See Section \ref{s:sine_wave} for more details.}
\vspace{0.5cm}
\label{f:sine_wave}
\end{figure*}

The setup we focus on in this section is a two-dimensional box of side length, density, and internal energy of unity, where the initial velocity field in each of the $x$- and $y$-directions is a sine wave with an amplitude of unity. We use $20$ cells in each dimension, and $100$ \VF tracers per cell, which are distributed uniformly as a Cartesian grid in the initial conditions; i.e.~each cell is sampled with a $10\times10$ grid of \VF tracers initially. As the system evolves, the fluid flows (in each dimension) from the region around a diverging point, where the velocity is zero and the velocity gradient is positive, into a converging point, where the velocity is zero and the velocity gradient is negative. \Fig{sine_wave} shows the state of the system at times $t=0.2$ (top and middle rows) and $t=0.35$ (bottom row) for three physically identical initial setups that are evolved in three technically different ways, as described below. The background colour in the top and bottom rows represents the ratio of the \VF tracer density to the fluid density, while the curves in the middle row give the one dimensional averaged profiles along the $x$-axis of the velocity in the $x$-direction (red), fluid density (green), and \VF tracer density (magenta).

\begin{figure}
\centering
\includegraphics[width=0.475\textwidth]{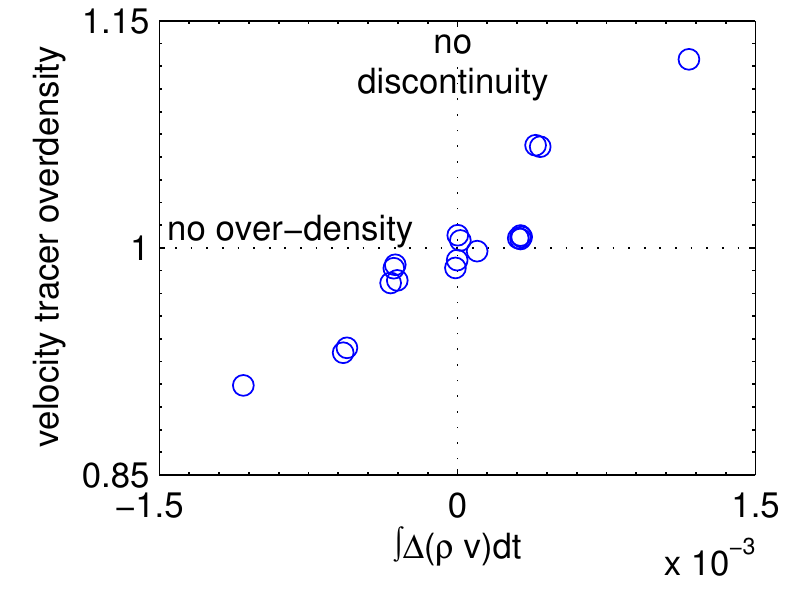}
\caption{The relation between \VF tracer over/under-density and the time-integrated local flux discontinuity across cell interfaces $\int{\Delta(\rho v){\rm d}t}$. Each data point represents the average of those quantities along the cell interfaces at one $x$-coordinate. The deviation of the \VF tracer density from the fluid density is built primarily at the locations of velocity discontinuities across cell interfaces. This occurs because the fluid is evolved according to the solution to the local Riemann problem, while the \VF tracers are advected according to the reconstructed velocity field and do not `see' the characteristics of the Riemann problem. The data points are measured at time $t=0.4$ from a setup similar to that shown in \Fig{sine_wave}, but where the amplitude of the initial velocity sine function is $0.1$ instead of $1$. With this smaller velocity, the over/under-densities are not advected away from the cell interfaces where they form, which makes the measurement cleaner.}
\vspace{0.5cm}
\label{f:discontinuities_tracer_densities}
\end{figure}

We begin with a discussion of the outcome shown in the middle column of \Fig{sine_wave}, where the mesh is static, and the diverging point lies at $x=0.25$, exactly along the interfaces between adjacent cells (this is true for $y=0.25$ as well; we hereafter refer only to coordinates along the $x$-direction, but the same holds for the $y$-direction due to the symmetry of the problem). It is immediately apparent from the $t=0.2$ panels that the density of the \VF tracers deviates significantly from that of the fluid in certain regions, in particular around the converging point, $x=0.75$. Strong deviations develop around certain cell interfaces, particularly those interfaces across which there exists a discontinuity in the reconstructed velocity field (red line in the middle panel). An example for such an interface is indicated with a dotted line in the middle panel at $x=0.7$. This is demonstrated explicitly in \Fig{discontinuities_tracer_densities} by showing a correlation between the time-integrated flux discontinuity across cell interfaces and the local \VF tracer over/under-densities at those interfaces. The middle bottom panel shows that once the \VF tracer over/under-densities form, they are advected with the fluid and do not diminish (unless, by chance, they come across a discontinuity that reverses the situation). This is seen in particular in the $x<0.5, y<0.5$ quartile, where a distorted Cartesian grid of \VF tracer over-densities appears. This is a result of the advection of the over-densities that developed along cell interfaces at $t\lesssim0.2$ in the $x>0.5, y>0.5$ quartile towards the new converging point $x=0.25$ at $t\gtrsim0.2$.

The reason for the development of \VF tracer over/under-densities is the following: the fluid is evolved according to the solution to the appropriate Riemann problem at the cell interface, while the \VF tracers do not obey that solution. Rather, the \VF tracers are advected according to the reconstructed velocity field. Where the velocity field is discontinuous in the direction of the velocity, the divergence of the field is non-zero, and the solution to the continuity equation gives a $\delta$-function at the location of the discontinuity. Two \VF tracers infinitesimally displaced from the interface, but to either side, will have arbitrarily large differences in their calculated velocities, and as a result their density at and around the interface can change dramatically, at odds with the solution to the Riemann problem. This occurs in our setup mainly around the converging point, $x>0.5$. When there is no discontinuity across an interface (generally at $x<0.5$ in this setup), the fluid and the \VF tracers are both simply advected across it. In such cases, the differences in the resulting effective fluxes between the fluid and the \VF tracers are typically small, and do not lead to $\delta$-function-like over/under-densities, however they are not identically zero. The reason is that the fluid and the \VF tracers are still not evolved using the same velocities; the fluid is advected with the reconstructed velocity at the interface (i.e~the Riemann solver only `sees' the values at the interfaces, which are indeed calculated using the gradients, but it does not explicitly `see' the gradients themselves), while the \VF tracers are advected with velocities interpolated to their positions inside the bulk of the cell. The differences arising across cell interfaces with no discontinuities can be estimated by considering the flux across such an interface during a single time step $\Delta t$, which is proportional to the maximal distance $l_m$ from the interface that fluid, or tracers, can reach (or arrive from) during $\Delta t$. Let us consider a linear velocity gradient $\frac{{\rm d}v}{{\rm d}x}$ around the interface, and define the interpolated velocity at a distance $l_m$ from the interface as $v_m\equiv v_i+\frac{{\rm d}v}{{\rm d}x}l_m$, where $v_i$ is the velocity of the fluid at the interface. Further we can write for the \VF tracers $v_m\Delta t\approx l_{m,{\rm tr}}$, and for the fluid $v_i\Delta t\approx l_{m,{\rm fluid}}$, since the Riemann solver does not directly see the gradient, and solve for $l_m$. A simple calculation gives that the flux of \VF tracers differs from that of the fluid by a factor $1-\frac{{\rm d}v}{{\rm d}x}\Delta t$. We parametrize $\frac{{\rm d}v}{{\rm d}x}\Delta x=\eta v_i$, where $\Delta x$ is the cell size, and note that usually $\eta\ll1$. We also write the Courant condition as $\Delta t=\xi\frac{\Delta x}{v_i}$, and note that $\xi\ll1$. Therefore, the flux of the fluid and that of the \VF tracers in one time step differ by a factor of $1-\xi\eta$, where typically $\xi\eta\ll1$. Note that a higher-order predictor-corrector time advancement scheme only changes the former analysis such that instead of the velocity at a distance $l_m$ from the interface, one has to consider the mean velocity between the initial position and the predicted final position. In this case, the flux difference factor becomes $\frac{1-0.5\xi\eta(1+\xi\eta)}{0.5(1+\xi\eta)}$; i.e.~it is of the same order of magnitude as for the Euler time advancement scheme. Discussions along the same lines can be found in \citet{PopovP_08a} and \citet{McDermottR_08a}, where more sophisticated reconstruction methods are implemented, as well as a sub-timestep advection scheme at velocity discontinuities. It is found that those methods can improve the accuracy of their \VF tracer scheme (however, their reconstruction method is limited for Cartesian grids). Nevertheless, the scheme still suffers from conceptual problems that limit its accuracy in principle, as discussed below.

The left and right columns of \Fig{sine_wave} demonstrate that the evolution of the \VF tracers is in fact not independent of the fine details of the mesh configuration and motion. This lack of invariance to numerical parameters related to the volume discretization arises due to the dependence of the reconstructed velocity field on the mesh. In the left panels we show the result from a setup physically identical to the one in the middle column, but where the initial conditions were displaced by half a cell size with respect to the mesh, such that the converging point lies at the centre of the cell instead of on the cell interface. The resulting deviations of the \VF tracer density from the fluid density are different, as a result of a different pattern of discontinuities in the reconstructed velocity field. However, there is no fundamental difference between these two setups -- the physical and correct numerical solutions for the fluid are both unchanged. Indeed, in more complicated, less controlled, problems, such a difference in the initial conditions will bear no noticeable consequences. The point to be taken from this exercise is that the \VF tracer density deviations are directly related to the small-scale details of the interpolated velocities. A similar case is shown in the right column of \Fig{sine_wave}, where the initial conditions are the same as in the middle column, but the simulation uses a moving Voronoi mesh instead of a static Cartesian mesh. In this case, the mesh starts to be distorted around the converging point, losing its Cartesian nature due to the mesh regularization procedure, which works to keep the cells roughly `round' (note, however, that that is not the case at the location where both axes converge, due to the symmetry there). As a result, local, non-Cartesian, discontinuities arise between adjacent cells, and hence a complex pattern of the \VF tracer density field develops. The averaged one-dimensional profile is similar to those of the static mesh calculations, but the local \VF tracer over/under-densities are markedly different due to the additional features of the velocity field that develop where the mesh is distorted. Hence, in this rather simple symmetric flow, the distortions that result from the regularization of the moving mesh lead to a somewhat worse performance of the \VF tracers. As shown below in Section \ref{s:turbulence}, a fixed Cartesian mesh does not have an advantage for a more complex flow that does not have special directions that parallel the mesh. We also note that with increased resolution the deviations become smaller, as the velocity field can be reconstructed more continuously. This is true in general in problems where the smallest scales are fixed such that higher resolution resolves them better. However, in the more complex setups presented below in Section \ref{s:complex_tests}, as the resolution is increased, there is also additional power on small scales, such that there always exist discontinuities in the reconstruction. In such cases, the increased resolution is less effective in improving the performance of the \VF tracers.

To avoid these adverse effects of discontinuities, an interpolation scheme that provides a continuous interpolated \VF should be used. This can be achieved by, e.g.~a tri-linear interpolation using neighbouring cells, or higher order schemes. However, such schemes do not preserve the mean velocity within each cell; i.e.~the mean of the interpolated \VF inside a given cell is not guaranteed to be equal to the mean fluid velocity of the cell (e.g.~\citealp{McDermottR_08a}). This means that the mass flux represented by the \VF tracers will in general differ from the mass flux of the fluid itself. Instead, to ensure that the tracers follow the fluxes of the fluid, the full solution to the Riemann problem would have to be constructed, and the \VF tracers then advected according to the \VF given by that solution. Complications such as determining which Riemann problem each \VF tracer `sees' (as each cell has multiple faces) would come up, and if dealt with appropriately, a situation can possibly be achieved where if the \VF tracer density field matches the (reconstructed) density field of the fluid at the beginning of a time step, the correct number of \VF tracers will move across the cell interface during a single time step. \citet{ZhangY_04a} introduced an alternative approach, where consistency between the tracer flux and fluid flux is achieved using `correction velocities' calculated especially for this purpose. However, there remains a fundamental problem with the \VF tracer approach that will nevertheless not allow the \VF tracers to follow the fluid correctly.

Recall that each time step in \AREPO{ }can be considered as a sequence of three steps: reconstruction, evolution, and averaging \citep{SpringelV_10a}. So far our discussion had to do with the differences between the fluid and the \VF tracers in the evolution step. However, the reconstruction and averaging steps are completely absent for the \VF tracers. This means not only that the over/under-densities are advected and do not diminish once formed, as already discussed, but also that even if a scheme was devised that avoided (or corrected) the problem in the evolution step, the distribution of \VF tracers at the end of their evolution step would in general differ from the distribution of the fluid after its averaging and reconstruction steps. Hence, the next evolution step will also differ between the fluid and the \VF tracers, as the initial condition for that step would already be different between the two components. To overcome this fundamental problem, the position of the \VF tracers would have to be made uniform, or randomized, inside each cell, imitating the averaging step, and then changed according to the reconstruction of the fluid density field. By doing that, however, the trait of the \VF tracers as having their own positions would be lost\footnote{\citet{MuradogluM_01a} and \citet{ZhangY_04a} discuss velocity and position correction algorithms that enforce consistency between the particles and the fluid `by hand'. However, those corrections indeed introduce diffusion, and can be quite aggresive, especially for low particle numbers, when they attempt to correct for the `shot noise' of the fluxes of individual particles.}. In fact, by combining the two `solutions' (that to the evolution step, and that to the averaging+reconstruction steps), the \VF tracers would effectively behave exactly like the \MC tracers, but with significant overhead of complication and computing time. Therefore, these `solutions' are in fact no solutions at all, as they imply a fundamental modification to the method. In other words, passive tracers that are advected with the local velocity field cannot be made to follow the mass flow of the fluid.

\section{Astrophysically-motivated tests: results}
\label{s:complex_tests}
\subsection{Driven isothermal turbulence}
\label{s:turbulence}
In this section we repeat several of the driven subsonic isothermal turbulence experiments of \citet{BauerA_12a}, while adding both types of tracer particles. We are motivated to study the performance of our tracer particle schemes in this problem, as we may expect it to be a challenging test based on our findings in previous sections, in particular for \VF tracers. A second motivation for the study in this section is that \VF tracers are commonly used in the literature to study mixing and various statistical properties of turbulence (e.g.~\citealp{YeungP_02a,BiferaleL_04a,BiferaleL_05a,ArneodoA_08a,FederrathC_08a,KonstandinL_12a}). Finally, turbulence appears in various contexts in astrophysics, and subsonic turbulence in particular may play a key role in the evolution of the intracluster and intergalactic media \citep{VogelsbergerM_12a,KeresD_12a}.

\begin{figure*}
\centering
\subfigure[]{
          \label{f:turb_density_maps_unsmoothed}
          \includegraphics[width=1.0\textwidth]{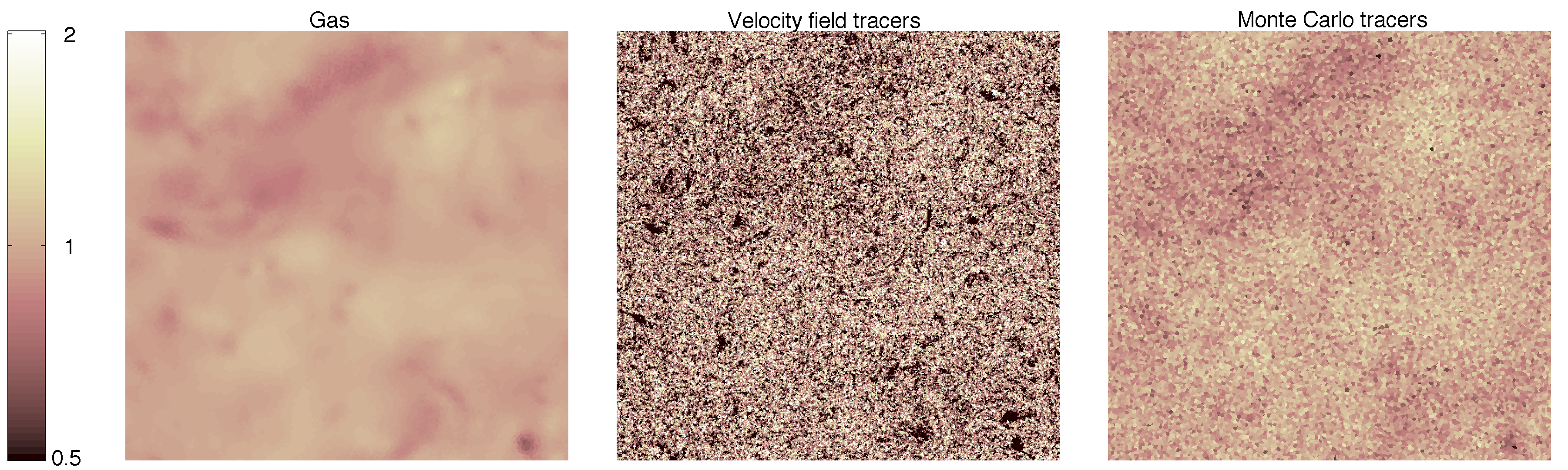}}
\subfigure[]{
          \label{f:turb_density_maps_smoothed}
          \includegraphics[width=1.0\textwidth]{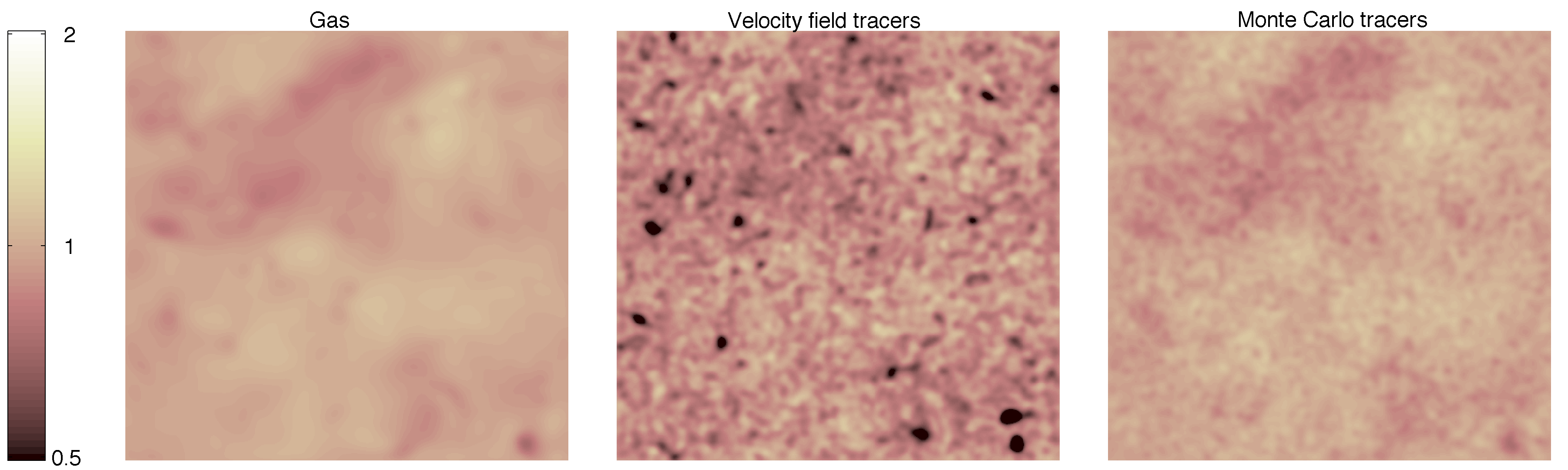}}
\caption{Thin slices through the density field of the gas (left), \VF tracers (middle), and \MC tracers (right), for the $128^3$ turbulent box with $64$ tracer particles per cell, at time $t=25$. The images do not take cell gradients into account, since self-consistent gradients cannot be constructed for the tracer particles. In the top panels, which show the unsmoothed density field, it can be seen that both tracer species reproduce the perturbations in the density field of the gas on the largest scales, but on small scales they behave differently. In particular, the \VF tracers display structures that do not exist in the gas distribution, while the \MC tracers introduce excess small-scale power due to Poisson noise. The \VF tracers show very disparate density values from the mean density, partly because their positions do not undergo any mesh-regularization procedure as opposed to the fluid. To isolate this effect, we show in the bottom panels the same fields but smoothed with a two-dimensional Gaussian with a width of $1/128$; i.e.~four times the mean \VF tracer particle separation, which is large enough to smooth out any effects of mesh irregularity. Indeed, the densities in the smoothed version are much closer to the mean density, nevertheless strong deviations from the density field of the fluid still exist. These are attributed solely to the non-Lagrangian nature of the \VF tracers as discussed in the text.}
\vspace{0.5cm}
\label{f:turb_density_maps}
\end{figure*}

\begin{figure*}
\centering
\subfigure[]{
          \label{f:turb_density_PDF}
          \includegraphics[width=0.475\textwidth]{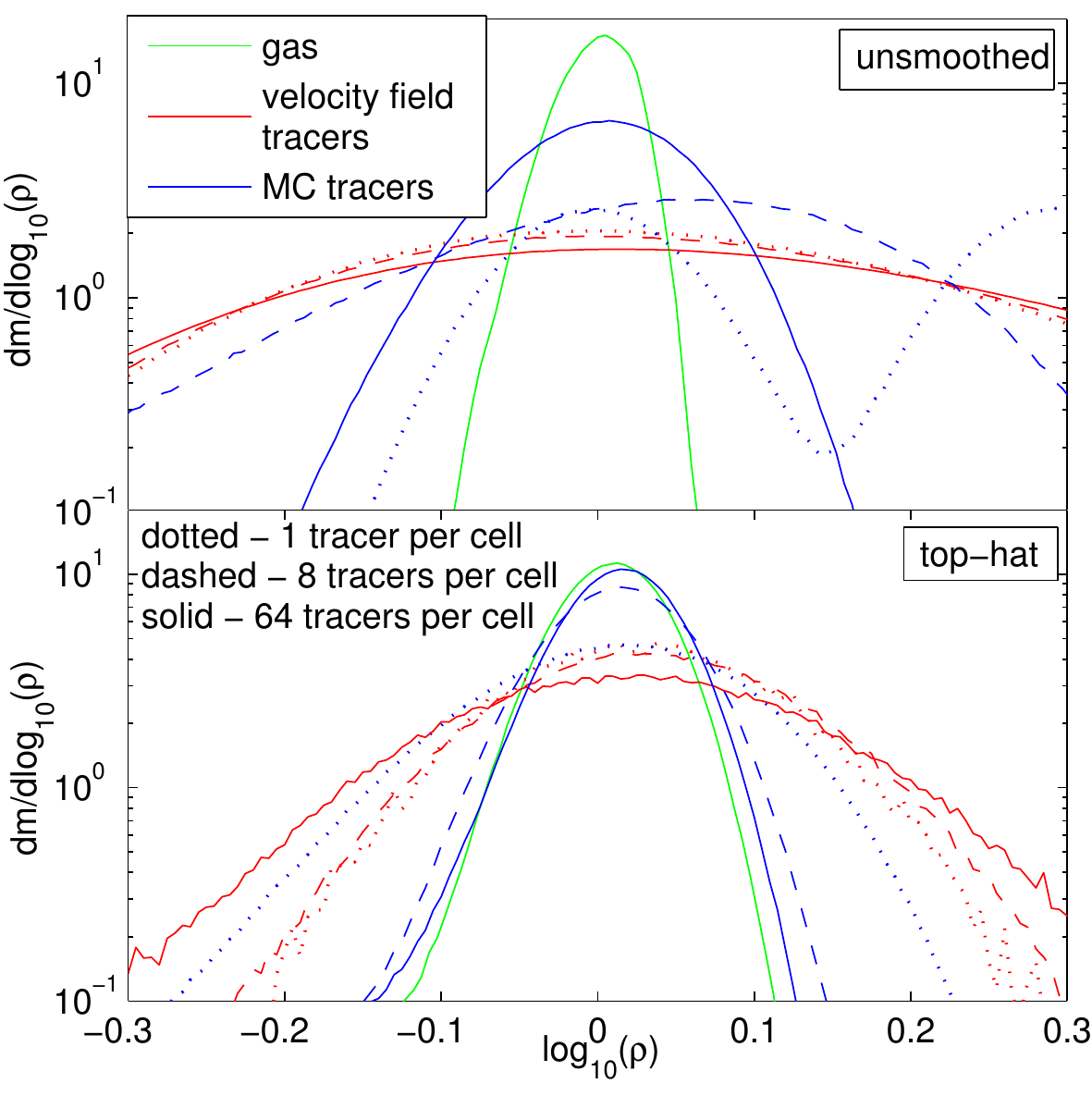}}
\subfigure[]{
          \label{f:turb_density_power_spectra}
          \includegraphics[width=0.475\textwidth]{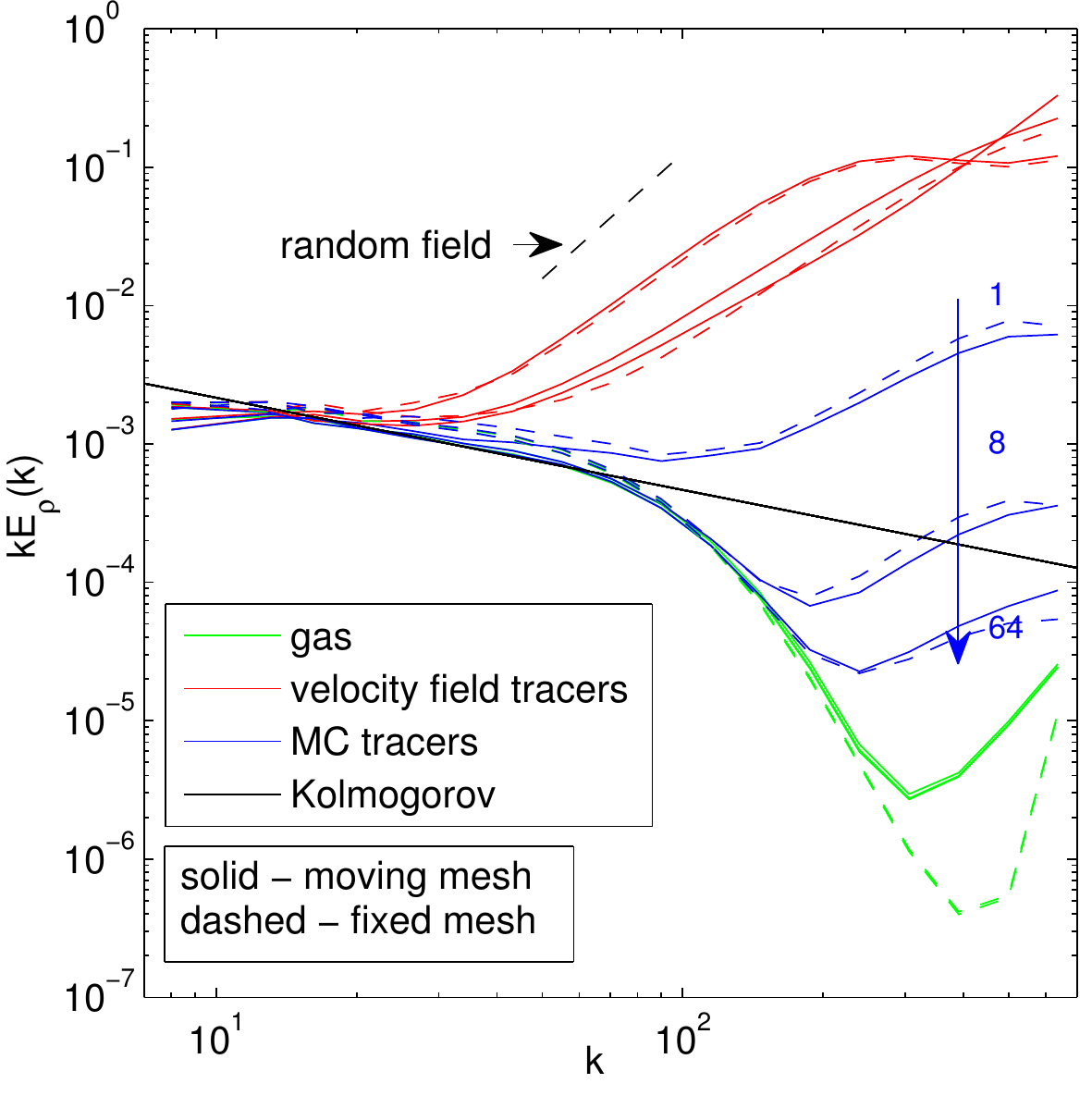}}
\caption{{\it Left:} Mass-weighted density PDFs of gas and tracers in a $128^3$ isothermal driven subsonic turbulence test at the final time $t_f=25.5$. The density PDF of the \VF tracers (red) is independent of how many tracers are used (different dashed curves), and follows quite poorly the PDF of the gas (green), being much broader. For the \MC tracers, discreteness is clearly seen when a small number of tracers per cell $\NMC$ is used. For example, with one \MC tracer per cell, a second peak appears around twice the mean density of the gas (dotted blue) -- these are cells with two \MC tracers. However, as $\NMC$ is increased, the discreteness is not seen anymore, and instead the PDF is broader than the gas PDF by $\approx{\rm log}_{10}(1+\NMC^{-0.5})$. In the bottom part of the panel, all the densities are smoothed with a Top-Hat kernel of 32 neighbours. In this case as well, the \VF tracer density PDF deviates from that of the gas, indicating that mesh-regularization issues are not the cause for the mismatch (see text). {\it Right:} Density power spectra for the same problem, averaged over $t=20-25.5$, and shown up to a wave-number $k$ approximately corresponding to the mean particle separation. The power spectrum of the gas (green) follows the expected Kolmogorov power spectrum until some small scale where dissipation kicks in. The power spectrum of the \VF tracers has a different slope, with much more power on small scales. The power spectra of the \MC tracers converge towards that of the gas as $\NMC$ is increased, regardless of whether the mesh is moving (solid) or static (dashed). We note that at different resolutions (not shown), different maximum wave-numbers are resolved, however the relative shapes of the tracers power spectra compared with the gas are unchanged.}
\vspace{0.5cm}
\label{f:turb_density_stats}
\end{figure*}

We run boxes of side length unity with a total mass of unity, where the gas is kept isothermal with unit sound speed. The gas is distributed uniformly with zero velocity in the initial conditions at $t=0$, when the turbulence driving is turned on. The boxes reach a quasi-steady state at time $t\approx5-10$ and are evolved until time $t_f=25.5$ (for full details on the turbulence driving force and the measurement of power spectra, we refer the reader to \citet{BauerA_12a}). We examine subsonic turbulence with a Mach number $M\approx0.3$. We run the box at four resolution levels, with $32^3$, $64^3$, $128^3$, and $256^3$ cells, and use both a static and a moving mesh. In addition, we vary the number of tracer particles, using $1$, $8$ and $64$ tracers per cell, of both tracer species. In \Fig{turb_density_maps} we show thin slices through the density field of the gas (left), \VF tracers (middle), and \MC tracers (right), for the $128^3$ box with $64$ tracer particles per cell. In \Fig{turb_density_stats} we show the density probability distribution functions (PDF; top) and the density power spectra (bottom) measured from our $128^3$ runs.

The complexity of the flow in compressible isothermal turbulence results in large differences between the evolution of the fluid based on the Riemann solver (and the subsequent averaging and reconstruction steps) and the evolution of the \VF tracers based on the interpolated velocity field. Therefore, it is not surprising that spurious features develop in the projected density of \VF tracers, as can be seen in the middle panels of \Fig{turb_density_maps}. Qualitatively similar images are shown by \citet{PriceD_10a} (their Fig.~5) for supersonic turbulence simulated using the {\small FLASH} code. There, the interpolation of the \VF is performed with a Cloud-In-Cell scheme that does not produce discontinuities, yet the \VF tracers are strongly biased with respect to the fluid, in line with our results and the discussion in Section \ref{s:sine_wave}. Similarly to the case shown in the bottom panels of \Fig{sine_wave}, \VF tracer over/under-densities (with respect to the gas) are not necessarily correlated with over/under-densities of the gas itself. This is because the former are advected for some time after they form, a time that can be longer than the time it takes the gas to undergo significant compression/rarefaction. \Fig{turb_density_stats} demonstrates clearly that the \VF tracers do not accurately represent the compressible fluid in a turbulent flow. In \Fig{turb_density_PDF}, the density PDF of the \VF tracers (red) is shown to be significantly broader than that of the gas (in agreement with \citealp{PriceD_10a}). The systematic difference is even more dramatic for the density power spectra, shown in \Fig{turb_density_power_spectra}, as the \VF tracers are strongly correlated on small spatial scales. The shape of the density power spectrum of the \VF tracers is completely erroneous, and does not converge with increasing number of tracers per cell, since it is not the result of random sampling noise but represents a truly systematic inability of the method to recover the underlying gas motions.

In contrast, the \MC tracers are able to follow the density power spectrum of the gas $E_{\rho}(k)\propto k^{-5/3}$ very accurately, until the random sampling noise kicks in at the smallest scales and gives their power spectrum a slope close to that of a random field, $E_{\rho}(k)\propto k$. The length scale at which the random noise takes over depends on the number of \MC tracers per cell, as shown in \Fig{turb_density_power_spectra}. This random noise can also be seen in the deviations of the density PDF in \Fig{turb_density_PDF}, which is broader for the \MC tracers than for the gas. However, with a high enough number of tracers per cell, the density PDF of the \MC tracers can reproduce that of the gas to arbitrary accuracy. The second moment of the gas density PDF is $\approx0.02$, while for the \MC tracers the second moments are $\approx0.18$ and $\approx0.06$ when $8$ and $64$ tracers per cell are used, respectively. This corresponds to the expected broadening by Poisson noise, as ${\rm log}_{10}(1+\sqrt{8})\approx0.13$ and ${\rm log}_{10}(1+\sqrt{64})\approx0.05$. As expected, there is a very small dependence on whether the mesh is moving or not, as the \MC tracers follow the mass by construction regardless of that property of the mesh.

It is worth pointing out a technical point regarding Figs.~\ref{f:turb_density_PDF} and \ref{f:turb_density_maps_unsmoothed}. The densities of the \VF tracers that are used to plot the quantities in those figures are derived by constructing a Voronoi mesh with the \VF tracers as the mesh-generating points. However, since no mesh regularization procedure is applied to those points, the cell volumes are expected to be irregular. This introduces some scatter in the densities, which will broaden the density PDF and generate power on the smallest scales, of order the cell size. The bottom panel of \Fig{turb_density_PDF} shows that that is not the determining factor in the differences between the statistics of the \VF tracers and the fluid. This is shown by using a $32$-neighbour Top-Hat smoothed density field, for all components. In this case, small-scale mesh regularity does not affect the densities, however, the PDF of the \VF tracers is still very different from that of the gas, and is not converging towards it. In contrast, the PDF of the \MC tracers is much closer to that of the gas, simply because with the smoothing procedure the effective $\NMC$ is larger. Similarly, \Fig{turb_density_maps_smoothed} shows the density fields from \Fig{turb_density_maps_unsmoothed} after applying smoothing on the scale of four times the mean \VF tracer particle separation, such that mesh regularization issues are irrelevant. That makes it easier to see the correspondence to the fluid on the largest scales, and at the same time the strong deviations that exist.

\begin{figure}
\centering
\includegraphics[width=0.475\textwidth]{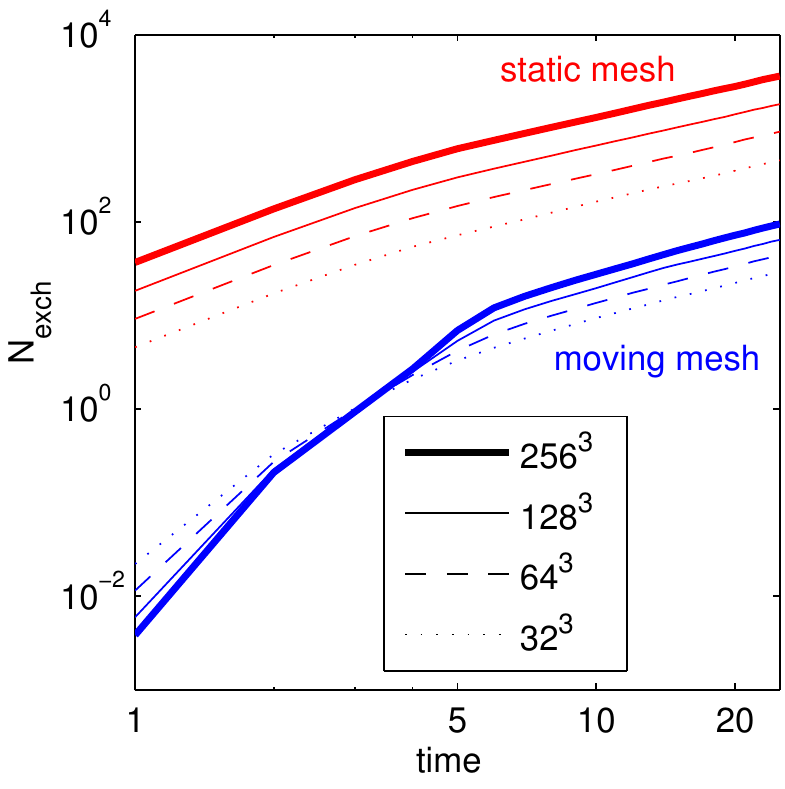}
\caption{The number of exchanges between cells, $N_{\rm exch}$, that \MC tracers experience as a function of time in the various turbulent box runs. Both calculations with a moving and a static mesh are shown, for different resolution levels, as indicated by the legend. $N_{\rm exch}$ is dramatically reduced in the moving mesh calculations with respect to those with a static mesh, and its dependence on the resolution is weaker, both indicating that numerical mixing is reduced thanks to the quasi-Lagrangian nature of the \AREPO{ }moving mesh.}
\vspace{0.5cm}
\label{f:turb_Nexch}
\end{figure}

We end this section with an investigation of the effect of the movement of the mesh on the mixing of the fluid, which can be quantified with the \MC tracers. To this end, we use the number of cell exchanges that each \MC tracer undergoes, $N_{\rm exch}$, as a probe for the amount of mixing, since each such exchange is associated with transfer of fluid from one cell to another, and mixing within that new cell of the `old' and `new' fluids. In \Fig{turb_Nexch} we show the average number of cell exchanges (per \MC tracer) as a function of time for various runs, namely at four resolution levels, $32^3$ to $256^3$, and for both static and moving mesh simulations. We identify several interesting trends.

First, a common feature of all the curves is an early rapid growth of $N_{\rm exch}$ at $t\lesssim10$ when the turbulent cascade is built up, and later a quasi-steady state is seen with a linear growth of $N_{\rm exch}$ with time. Second, very large differences exist between the calculations with a static and with a moving mesh, as $N_{\rm exch}$ is approximately two orders of magnitude larger in the simulations with a static mesh. Third, for static mesh calculations, there is an inversely-linear relationship with the spatial resolution, $N_{\rm exch}\propto(\Delta x)^{-1}$, at all times. This can be understood if one thinks of the trajectory of gas elements (represented by the tracers) as unchanged with resolution, which is a good approximation since most of the power is on the largest scales. Then, a larger number of cells have to be crossed as the cells become smaller, for the same trajectory. Fourth, the scaling of $N_{\rm exch}$ with resolution in moving mesh calculations is more complex. At early times, $t\lesssim10$, when the flow is simple as the turbulent cascade has not developed yet, we find that $N_{\rm exch}\propto(\Delta x)^{0.9}$; i.e.~a reverse trend is seen -- the higher the resolution, fewer cell exchanges are required. This is because the moving mesh is able to follow the (simple) flow very well, thereby minimizing fluxes between cells. However, at late times when the turbulence is fully developed, we find that $N_{\rm exch}\propto(\Delta x)^{-0.55}$; i.e.~more exchanges are needed in better resolved flows. This means that the mesh motion is unable to fully account for the (complex) flow anymore, and mass fluxes between cells are required. Nevertheless, the relationship is not inversely-linear as in the static mesh case, but weaker than that. This shows that the moving mesh is able to reduce the fluxes between cells by partially accounting for the flow with its own motion. Both the reduced magnitude and the weaker dependence of $N_{\rm exch}$ on resolution are indications for reduced numerical mixing achieved in the moving mesh calculations compared to ones with a static mesh.

\subsection{Cosmological simulations}
\label{s:cosmo}
In this section we investigate the performance of the tracer particle schemes in a cosmological context, by running cosmological simulations of structure formation with both tracer species included. The periodic box has a comoving volume of $(20\hMpc)^3$, and is evolved in two resolution levels: one using $128^3$ dark matter collisionless particles and $128^3$ initial gas cells, and one with $256^3$ resolution elements of each kind. The initial conditions are the same as in \citet{VogelsbergerM_12a}. In addition, the initial conditions include one \VF tracer particle per cell and one \MC tracer particle per cell. Both runs are non-radiative, such that gas does not cool and does not form stars. This is done so that we can compare the two tracer schemes on the same footing, otherwise differences would appear due to the fact that when gas turns into stars, \MC tracers can be associated with the new star particle, while the \VF tracers would be `left behind' and continue being advected with the gas.

\begin{figure*}
\centering
\includegraphics[width=1.0\textwidth]{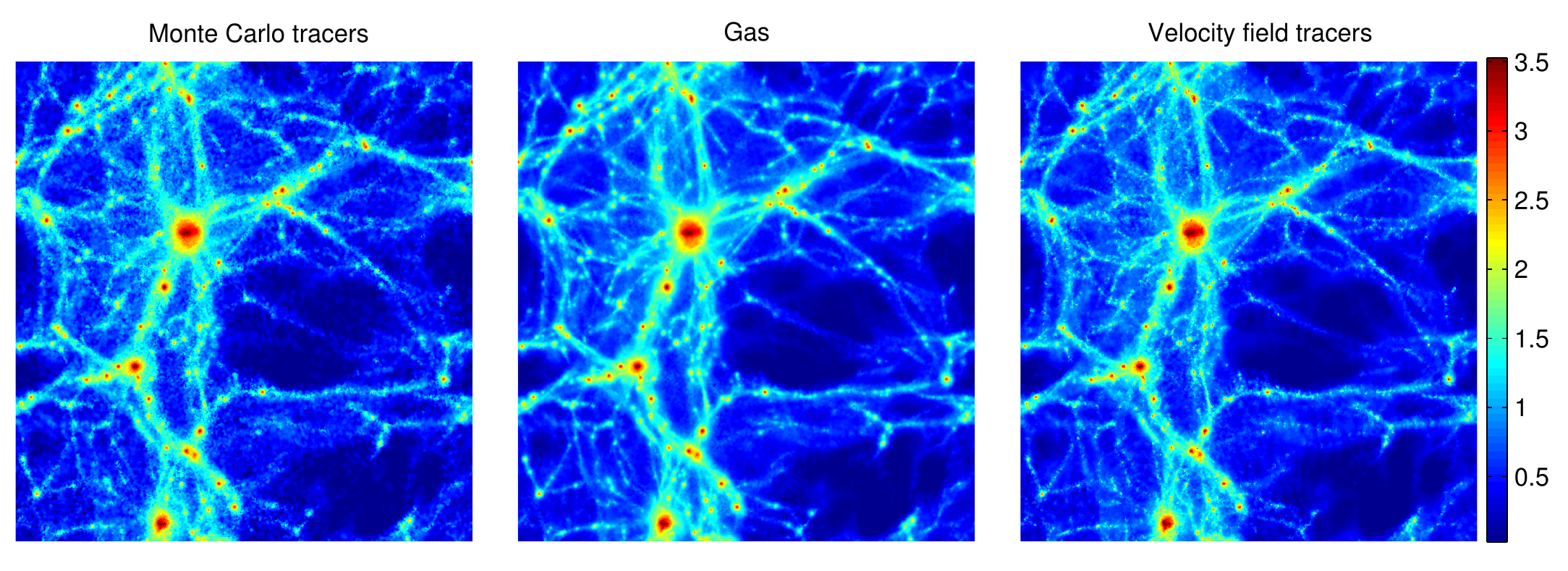}
\caption{Density projections of a non-radiative $128^3$ cosmological simulation at $z=0$. The side length of the images is $20\hMpc$ (the full extent of the simulation box), and the density is shown for a slice $5\hMpc$ thick around the centre of the box. The values indicated on the colour bar are in units of $\log_{10}(h^2\Msun\kpc^{-3})$. On first inspection, on the large scales, the distribution of the \MC tracers (left), as well as the \VF tracers (right), follows the gas distribution (centre) very closely. However, a closer examination of small scale details reveals some deviations. The appearance of the \MC tracers image is more `granular' as a result of \MC sampling noise, an effect that can be seen at all densities. The \VF tracer density field also appears `granular', but only in high density regions (haloes and filaments), and for a different reason. In those regions of active structure formation, the hydrodynamical flow is complex and violent, which causes strong local biases in the density of the \VF tracers, as shown in Section \ref{s:sine_wave}. This makes the image look sharper and more clumpy. In addition, the filaments appear thinner, and halo centres more concentrated, an effect that is shown quantitatively in \Fig{cosmo_profiles}.}
\vspace{0.5cm}
\label{f:cosmo_maps}
\end{figure*}

In \Fig{cosmo_maps} we show density projections through a slice $5\hMpc$ thick across the whole simulation volume at $z=0$ (using the $128^3$ simulation), for gas (middle), \MC tracers (left), and \VF tracers (right). Clearly, large-scale structures are followed well by both tracer particle schemes, which demonstrates the gross reliability of the methods for such an application. However, the images appear different when small scales are examined more closely. The image based on the \MC tracers appears more `granular'. This is a result of the \MC sampling noise, which can make adjacent cells with similar gas densities have different \MC tracer densities. This effect occurs at all gas densities. The image based on the \VF tracers also appears more `granular' than the gas, but only in higher density regions such as filaments and haloes. In addition, filaments appear thinner, or sharper. These differences with respect to the appearance of the gas image originate from completely different reasons. For the \VF tracers, the reason for the `granularity', or `sharpness', is that in regions where the hydrodynamical history of the gas is complex, large over/under-densities of the \VF tracers are formed, as shown in Section \ref{s:sine_wave}. This occurs to a much lesser degree in smoother regions of the flow, as in the voids seen in \Fig{cosmo_maps}.

\begin{figure}
\centering
\includegraphics[width=0.475\textwidth]{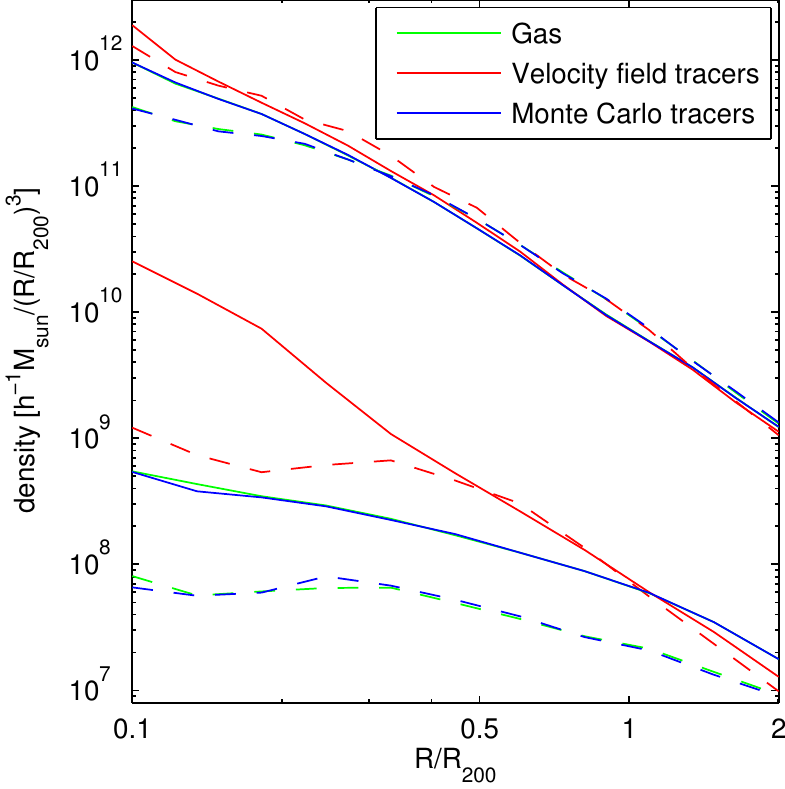}
\caption{Stacked gas and tracer density profiles of haloes taken from non-radiative cosmological simulations at $z=0$. The upper set of curves shows the profiles of around ten $M\sim10^{12}\Msun$ haloes, and the bottom set is for several hundred $M\sim10^{10}\Msun$ haloes. The dashed lines are from a $128^3$ simulation, where the $M\sim10^{10}\Msun$ haloes are barely resolved (with $\approx35$ resolution elements) and the solid lines from a $256^3$ simulation that has an eight times better mass resolution. A clear bias is seen in the \VF tracer profiles (red) with respect to the gas (green), while the \MC tracer profiles (blue) follow those of the gas. In particular, the \VF tracers are concentrated towards the centres of haloes, where their density can exceed that of the gas by up to about an order of magnitude. We interpret this effect as a consequence of the results presented in Section \ref{s:sine_wave}. Namely, gas in haloes in general, and at small radii inside haloes in particular, has a complex and violent dynamical history, during which \VF tracer concentrations develop that are never diminished, but rather fall with the most disturbed gas into the halo centre.}
\vspace{0.5cm}
\label{f:cosmo_profiles}
\end{figure}

A close investigation of the largest halo in the images will also show that its centre is more concentrated in the \VF tracer image compared to the gas. To quantify this effect, we measure the stacked density profiles of haloes at $z=0$ and show them in \Fig{cosmo_profiles}. We stack both haloes of mass $\sim10^{12}\Msun$ (top curves) and $\sim10^{10}\Msun$ (bottom curves), and show results from the $128^3$ simulation (dashed) and the $256^3$ one (solid). \Fig{cosmo_profiles} demonstrates that the deviations of the flow of the \VF tracers from that of the gas, which originate from sub-cell differences as shown in Section \ref{s:sine_wave}, have a profound effect in the cosmological context. The \VF tracers tend to concentrate at the centres of haloes, where the density they represent can exceed that of the gas by even an order of magnitude. For the barely-resolved $M\sim10^{10}\Msun$ haloes in the $128^3$ simulation, this concentration effect is even seen at all $R\lesssim R_{200}$, which shows clearly that not only the internal distribution of the \VF tracers in halos is biased compared to the gas, but the halo mass function itself is biased (overestimated) when the tracers are used to measure baryonic mass. This is not a random effect, but a clear systematic bias. We interpret this bias as a result of \VF tracer over-densities that develop when the gas that later ends up in the inner regions of haloes is part of a complex flow where discontinuities in the reconstructed velocity field are common and strong, and hence the deviations between the reconstructed velocity field and the solutions to the local Riemann problems between cells are the largest. The effect is particularly noticeable at low resolution; i.e.~in lower mass halos and for the lower resolution simulation. This is because the velocity field reconstruction is smoother, i.e.~less discontinuous, when a given system is simulated with more resolution elements.

The results shown in \Fig{cosmo_profiles} imply that \VF tracers fail to correctly follow gas flows in cosmological simulations. The very fact that the density of \VF tracers towards the centres of haloes is much larger than the gas density means that the tracers there do not have the same accretion history as the gas. Hence, following their trajectory and (thermo-)dynamical history in the Lagrangian sense does not provide a reliable measure of the same quantities for the gas. Therefore, in our study of galactic gas accretion \citep{NelsonD_13a} we use the \MC tracer particle approach in our analysis, having demonstrated their superior reliability in this particular application.

\section{Application: Thermodynamical history of the Santa Barbara Cluster atmosphere}
\label{s:cluster}
In this section we perform a simple analysis of an astrophysical setup, namely the `Santa Barbara Cluster' \citep{FrenkC_99a}, using the \MC tracer particles, in order to demonstrate the new capabilities that they add to \AREPO. We use the standard initial conditions of the cluster with $128^3$ resolution elements for each of the dark matter and gas components, and in addition initially assign $8$ \MC tracer particles to each gas cell\footnote{We also included the \VF tracers in the simulation, and verified that they spuriously accumulate in the centre of the cluster, in a similar way to the results shown in Section \ref{s:cosmo} and \Fig{cosmo_profiles}. We therefore do not consider them for the analysis in this section.}. The simulation is run in non-radiative mode; i.e.~the only processes taken into account are gravity and hydrodynamics for an ideal gas. The tracers record the maximum temperature along their evolution and the time they reached that temperature, as well as the maximum entropy and its associated time, as described in Section \ref{s:recording}.

\begin{figure*}
\centering
\includegraphics[width=1.0\textwidth]{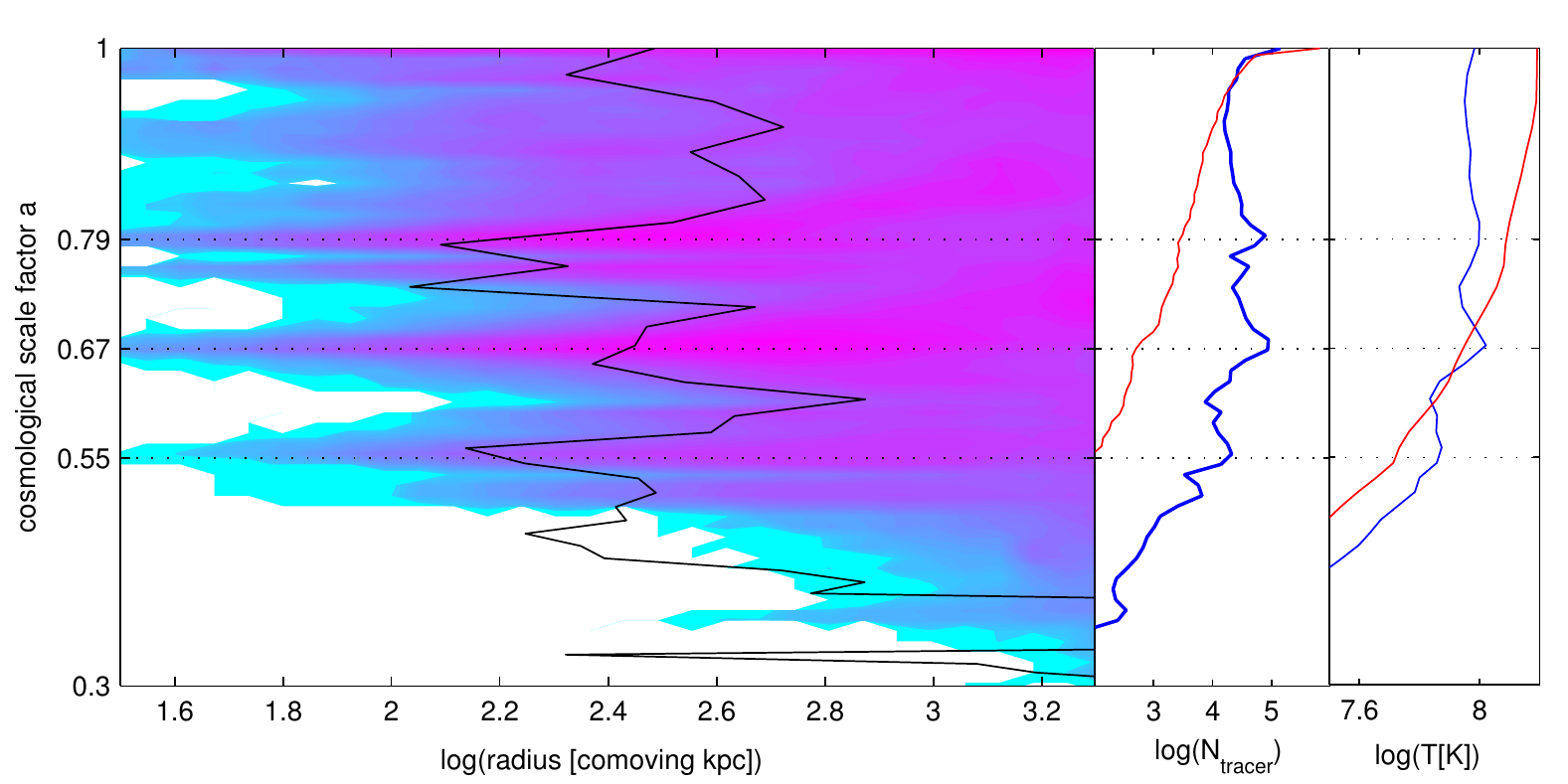}
\caption{{\it Left}: A two-dimensional histogram of the distances of \MC tracers from the Santa Barbara Cluster center-of-mass at $z=0$ and the cosmological scale factor at which they recorded their maximum past temperature (background colour). The displayed radius range corresponds roughly to the virial radius at $z=0$, $R_{200}\approx2.3\Mpc$. The signature of the dynamical evolution of the cluster, characterized here by the distance of the most massive satellite from the cluster center (black), can be clearly seen in the thermal history of the gas, as times when a substantial fraction of the gas, at a large range of radii, experiences its peak temperature. {\it Middle}: One-dimensional histograms of the times at which the past temperature (blue) and past entropy (red) are at their maximal values. The maximum past temperature time of a significant fraction of the mass is well before $z=0$, while the maximum past entropy time is peaked strongly at $z=0$. {\it Right}: Instantaneous mean mass-weighted temperature inside the main progenitor of the cluster (blue), and the mean mass-weighted cluster entropy (red; on an arbitrary scale), shown as a function of the scale factor. It can be seen that the dynamical events that heat the gas (marked with dashed horizontal lines) are followed by (adiabatic) cooling, while the entropy, as expected for a non-radiative simulation, is always increasing.}
\vspace{0.5cm}
\label{f:SB_timeTmax_vs_R}
\end{figure*}

\Fig{SB_timeTmax_vs_R} shows the distribution of \MC tracers in the two-dimensional plane of distance (at $z=0$) from the cluster center-of-mass and the time (denoted by the cosmological scale factor $a$) at which the maximum past temperature was reached. The physical scale that is shown corresponds roughly to the virial radius of the cluster at $z=0$. An interesting pattern appears, where there exist special times ($a\approx0.55,0.67,0.79$) when the gas that is at small radii deep inside the cluster heats up, each followed by an outgoing pattern of gas reaching its maximum past temperature at larger radii as time progresses\footnote{We find that plotting on the $x$-axis the distance from the cluster center {\it at the time the maximum past temperature is reached}, instead of the distance at $z=0$, makes no qualitative difference to the appearance of the plot.}. This suggests that the cluster undergoes violent dynamical events at those times. To check this hypothesis, we plot (black) the distance of the second-most massive SUBFIND subhalo \citep{SpringelV_01} from the center of the cluster (i.e.~the most massive subhalo after the main halo itself). Indeed, there is a strong correspondence between the cluster-centric radius of the most massive satellite and the inner-most radius where significant heating occurs.

It is noteworthy that a significant amount of gas that is heated by these dynamical events does not surpass the associated peak temperature at later times, all the way down to $z=0$. This can be clearly seen in the collapsed one-dimensional distribution of maximum past temperature times that is shown in the middle panel of \Fig{SB_timeTmax_vs_R}, which features multiple peaks (blue). In fact, the gas cools (adiabatically) to temperatures below the maximum reached during those events, as demonstrated (blue) in the right panel of \Fig{SB_timeTmax_vs_R}, which shows the instantaneous mean gas temperature as a function of the scale factor. What the tracer analysis uniquely allows to determine is that in each of these violent close passages, a different subset (at least partially) of the cluster mass is heated. Otherwise, there would not exist multiple peaks in the distributions shown in the left and middle panels of \Fig{SB_timeTmax_vs_R}.

In contrast with the temperature evolution, the distribution of the maximum past entropy times (middle panel, red) is strongly peaked close to $z=0$. This indicates, as expected for a non-radiative simulation, that the entropy generated during the dynamical events in which the gas reaches its maximum past temperature does, for the most part, not decrease after those events as the gas adiabatically cools, but continues to gradually increase. This is further demonstrated by monotonicity of the instantaneous mean entropy shown in the right panel (red).

\section{Summary}
\label{s:summary}
In this paper we have presented the implementations and detailed characterization of two independent methods that attempt to follow the flow of the fluid in the hydrodynamical moving-mesh code \AREPO. The use of special methods is necessary for grid-based codes in order to follow the fluid in a Lagrangian manner, due to mass exchange between cells and their Eulerian nature. With such a procedure, the thermodynamical histories of fluid parcels can be followed, which can be important for certain astrophysical, as well as other, applications. For example, we used our methods to follow the cosmological gas accretion from the intergalactic medium onto dark matter halos and galaxies, and found that `hot mode' accretion is the dominant channel providing large galaxies at $z=2$ with their gas \citep{NelsonD_13a}.

The first method we presented is `\VF tracers', where passive particles are advected with the local interpolated velocity field given by the sub-grid reconstruction of the fluid velocity. This method has been used in the past in astrophysical AMR codes. We have shown, in agreement with previous work outside the astrophysical community, that this method cannot in fact follow the fluid correctly. All three steps that comprise together one \AREPO{ }time step, namely reconstruction, evolution, and averaging\footnote{Most other AMR codes in cosmology can also be viewed as REA-schemes (reconstruct-evolve-average).}, are inherently different between the \VF tracers and the fluid. In the evolution step, the problem arises when the fluid is evolved according to the solution of the Riemann solver, to which the reconstructed velocity field serves only as the initial condition. However, the positions of the \VF tracers are updated without incorporating the solution to the local Riemann problems. This can possibly be improved upon, or even solved, however we have not pursued this direction, since the problems in the reconstruction and averaging steps are even more fundamental. In fact, these two steps are completely absent in the \VF tracers scheme, and introducing them into the scheme will change it in the most fundamental way, as the tracers will then no longer follow the local fluid velocity, but will be moved `randomly' inside their cells. The implications of those problems in astrophysically relevant contexts are severe. For example, we have shown that when the \VF tracers are used to follow a turbulent flow, they develop an entirely different density PDF and power spectrum than the fluid. In the cosmological context, we have shown that \VF tracers tend to concentrate at the centres of dark matter halos and produce large over-densities there. As a result, the history of the gas in the centres of halos cannot be followed reliably with this approach.

The second method we present, `\MC tracers', is a novel approach which follows the average mass flux correctly by construction, consequently incurring the expense of statistical noise. In this method, \MC tracers do not have their own phase-space coordinates, but instead belong to fluid (or other) mass elements, and are exchanged between them based on the mass fluxes between those elements as given by the physical processes that govern the simulation. As a result, the spatial distribution of the \MC tracers follows that of the mass, only with additional \MC noise. This noise, however, is of little concern, as the \MC tracers are computationally inexpensive, and can be used in large numbers. However, the method as we presented it is more diffusive than the fluid, an aspect that can be improved upon in future work. This disadvantage is however much reduced when a quasi-Lagrangian moving-mesh is used instead of a static mesh, since mass fluxes between cells are minimized. We demonstrated that the \MC tracers accurately reproduce the density field in various tests, including a cosmological simulation. The \MC approach also extends in a natural way beyond just fluids, as the tracers can be associated with collisionless particles, such as stars, black holes, and other sources or sinks of baryonic matter. In cosmological simulations, this allows the mass flow to be followed in its entirety, and provides a powerful analysis method for future state-of-the-art simulations that include complex flows of different baryonic phases.

\section*{Acknowledgements}
The simulations described in this paper were run on the Odyssey cluster supported by the FAS Science Division Research Computing Group at Harvard University. We thank Tom Abel, Andreas Bauer, Christoph Federrath, Ralf Klessen, Diego Mu\~noz, R{\"u}diger Pakmor, Paul Torrey, and Franco Vazza, for useful discussions. VS acknowledges support from the European Research Council under ERC-StG grant EXAGAL-308037.

\label{lastpage}

\end{document}